\documentclass[11pt,a4paper]{article}
\usepackage{longtable}
\usepackage{amsmath}
\usepackage{amssymb}
\usepackage{graphicx}
\usepackage{bm}
\usepackage{footmisc}
\usepackage{afterpage}
\usepackage{float}
\usepackage{lscape}
\usepackage{longtable}
\usepackage{float}
\usepackage{slashed}
\usepackage{morefloats}
\usepackage{afterpage}
\usepackage{color}
\usepackage{bbold}
\usepackage[square,comma,sort&compress,numbers]{natbib}

\newcommand*{\no}{\noindent}
\newcommand*{\bea}{\begin{eqnarray}}
\newcommand*{\eea}{\end{eqnarray}}
\newcommand*{\be}{\begin{equation}}
\newcommand*{\ee}{\end{equation}}
\newcommand*{\pd}{\partial}

\newcommand*{\pref}[1]{(\ref{#1})}

\newcommand*{\mn}{{\mu\nu}}

\newcommand*{\tr}{\mathrm{tr}}

\newcommand{\bma}{\begin{pmatrix}}
\newcommand{\ema}{\end{pmatrix}}

\title{Exploratory study of the off-shell properties of the weak vector bosons}

\author{Axel Maas, Sebastian Raubitzek, and Pascal T\"orek\\
Institute of Physics, NAWI Graz, University of Graz,\\
Universit\"atsplatz 5, A-8010 Graz, Austria}

\begin{document}

\maketitle

\begin{abstract}

Gauge invariance requires even in the weak interactions that physical, observable particles are described by gauge-invariant composite operators. Such operators have the same structure as those describing bound states, and consequently the physical versions of the $W^\pm$, the $Z$, and the Higgs should have some kind of substructure. To test this consequence, we use lattice gauge theory to study the physical weak vector bosons off-shell, especially their form factor and weak radius, and compare the results to the ones for the elementary particles. We find that the physical particles show substantial deviations from the structure of a point-like particle. At the same time the gauge-dependent elementary particles exhibit unphysical behavior.

\end{abstract}

\section{Introduction}

Physical states have to be gauge-invariant. This, almost trivial, statement applies also to the weak interaction \cite{Banks:1979fi,Frohlich:1980gj,Karsch:1996aw,Philipsen:1996af,Philipsen:1997rq,Laine:1997nq,Maas:2017wzi,Attard:2017sdn}. However, this has unexpected consequences: The elementary $W^\pm$-boson, the $Z$-boson and the Higgs, i.e., the states obtained from the fields in the Lagrangian, cannot describe physical observable states\footnote{That for the parameters in the standard model they decay is actually only a parametric effect, and has no implications for this \cite{Maas:2017wzi}.}, as they are not gauge-invariant beyond perturbation theory. Rather, equivalent gauge-invariant states have to be constructed using composite operators \cite{Frohlich:1980gj,Frohlich:1981yi,Karsch:1996aw,Philipsen:1996af,Philipsen:1997rq,Laine:1997nq,Maas:2017wzi}.

This is at first surprising, as a description of weak interactions in experiments using the elementary particles works out very well \cite{Bohm:2001yx,pdg}. This contradiction has been resolved by Fr\"ohlich, Morchio and Strocchi (FMS) \cite{Frohlich:1980gj,Frohlich:1981yi}. For this, they extended standard perturbation theory (PT). A double expansion of the gauge-invariant composite operators in both the Higgs vacuum expectation value and the couplings yields that the composite states have on-shell essentially identical properties as the gauge-dependent elementary degrees of freedom. This, at first sight very surprising, result has been confirmed in various lattice calculations \cite{Shrock:1985un,Shrock:1985ur,Maas:2012tj,Maas:2013aia}, and also extends to the rest of the standard model \cite{Frohlich:1980gj,Frohlich:1981yi}. For a review see \cite{Maas:2017wzi}.

However, it turns out that this is due to the quite special structure of the standard model. In more general theories this leads frequently to qualitative disagreement between the gauge-invariant states and the elementary degrees of freedom even on-shell \cite{Maas:2015gma,Maas:2017xzh}. This can again be treated with the same double expansion \cite{Maas:2017xzh}, which will be dubbed gauge-invariant perturbation theory (GIPT) here following \cite{Seiler:2015rwa}. Lattice results confirm both the qualitative disagreement and the adequacy of GIPT on-shell for such theories \cite{Maas:2016ngo,Maas:2018xxu,Lee:1985yi}. This is again reviewed in \cite{Maas:2017wzi}.

However, in absence of any discovery of new physics, these dramatic predictions cannot yet be tested experimentally. But it would still be good to test such a fundamental prediction of quantum gauge theories within the available physics, i.e., the standard model. This is indeed possible. While on-shell a description in both terms agree, off-shell this is no longer necessarily the case. This will be discussed in Section \ref{s:theo}. Such deviations need to be rather small, to be consistent with existing data. But if these deviations should be large enough to be observable in experiments they constitute an additional standard model background in new physics searches, which has not yet been accounted for.

In the present exploratory study we are primarily interested in the underlying mechanisms and orders of magnitude. We will therefore study only the reduced weak sector, i.e., the $W^\pm$, the $Z$, and the Higgs, and drop the remainder of the standard model. For this purpose, we use a, in principle, reliable method to calculate correlation functions, even of composite operators, in a manifest gauge-invariant way: lattice simulations. We describe our technical setup and details of these simulations in Section \ref{s:lattice}. Our primary observable is the form factor for the vector bosons probed by a weak vector current. The results are shown in Section \ref{s:ff}. From this quantity we also extract the weak radius of the vector bosons, which are consistent with an extended structure. We discuss a possible experimental setup to test this in Section \ref{ss:exp}. The form factor of the gauge-dependent particles, on the other hand, shows an unphysical behavior.

A secondary aim of this work is to test how well GIPT works off-shell, as it can also be applied in this case. This would be very useful, as an analytical tool to treat these composite operators off-shell would allow to include the remainder of the standard model and access time-like phenomena. Therefore, we compare the off-shell properties of the vector bosons in sections \ref{s:prop} and \ref{s:ff} to the GIPT predictions to a particular order in the double expansion. Our results show that off-shell substantial deviations can appear, though quantitatively they depend on the parameters. We contrast these results with the ones from PT, i.e., the ones obtained from the elementary, gauge-dependent degrees of freedom. The difference between both is a quantitative measure of the sub-leading terms in the vacuum expectation value of the Higgs.

From the results we conclude in Section \ref{s:con} that deviations between a manifest gauge-invariant description and the one using the elementary fields are, in principle, present and observable. However, given the exploratory nature of the present study, we would not take our numbers quantitatively serious yet. Still, as we also observe qualitative differences, our results also suggest that momentum-resolved form factors of the weak vector bosons are candidates for an experimental investigation and tests of both the underlying field theory as well as of GIPT as a tool.

For those who wish to skip the technical details of the lattice computation, we refer to the theoretical background in Section \ref{s:theo} as well as the results in Sections \ref{ss:prop}, \ref{ss:ff}, and \ref{ss:radius}, and a possible experimental test setup in Section \ref{ss:exp}.

\section{Theoretical background}\label{s:theo}

\subsection{Setup}

We consider an $\mathrm{SU}(2)$ gauge theory with a complex scalar doublet, i.e.,
\begin{align}
    \mathcal{L} &= -\frac{1}{4}W_{\mu\nu}^a W^{a\,\mu\nu} + \frac{1}{2}\tr\Big[\big(D_\mu X\big)^\dagger\,\big(D^\mu X\big)\Big] - \lambda\,\Big(\frac{1}{2}\tr\big[X^\dagger X\big]-v^2\Big)^2\;,
    \label{eq:lagrangian}
\end{align}
where $W_{\mu\nu}^a=\partial_\mu W_\mu^a - \partial_\nu W_\nu^a-g\,\epsilon^{abc}\,W_\mu^b W_\nu^c$ is the field-strength tensor with the gauge fields $W_\mu^a$ and the gauge coupling $g$, $D_\mu=\partial_\mu-\mathrm{i}\,g\,W_\mu^a\frac{\sigma^a}{2}$ is the covariant derivative with the Pauli matrices $\sigma^a$. $X$ is a matrix representation of the scalar $\phi$, i.e.,
\begin{align}
    X &= \begin{pmatrix}
             \phi_1 & -\phi_2^\dagger \\
             \phi_2 & \phantom{-}\phi_1^\dagger
         \end{pmatrix}\;,
    \label{eq:xmatrix}
\end{align}
where $\phi_i$ are the components of the usual complex scalar doublet $\phi$. 

Besides the gauge-symmetry there is a global $\mathrm{SU}(2)_\mathrm{c}$ symmetry acting only on the scalar, the custodial symmetry. The full symmetry transformations are
\begin{align}
    \frac{\sigma^a W_\mu^a(x)}{2}=W_\mu(x)  &\to G(x)W_\mu(x)G(x)^\dagger + \frac{\mathrm{i}}{g}\big(\partial_\mu G(x)\big)G(x)^\dagger\\
    X(x) &\to G(x)X(x)M^\dagger\;,
\end{align}
where $G(x)\in\mathrm{SU}(2)$ and $M\in\mathrm{SU}(2)_\mathrm{c}$. 

The classical potential allows for a Brout-Englert-Higgs (BEH) effect. The parameters will be chosen below to have this effect, and to have a phenomenology of the type encountered in the standard model. Thus, in a PT setting standard methods \cite{Bohm:2001yx} yield that the three $W$ receive a mass $m_{\textrm{W}}$ by absorbing three degrees of freedom of the scalar field, leaving only a single, massive (Higgs) scalar with mass $m_{\textrm{H}}$. Note, that because there is no QED, the three $W$ are degenerate, and the would-be $Z$ is the ordinary third component. We do not perform the usual base rotation to generate $W^\pm$ states \cite{Bohm:2001yx}, and thus our $W$ states remain pure weak eigenstates.

\subsection{The gauge-invariant weak vector boson}\label{s:propintro}

We now shift to a discussion of the physical, i.e., composite states. These need to be constructed such that they are non-perturbatively gauge-invariant, i.e., including the Gribov-Singer ambiguity, see \cite{Maas:2017wzi} for details.

A vector custodial triplet is a suitable gauge-invariant replacement for the gauge-dependent $W$ bosons \cite{Frohlich:1980gj,Frohlich:1981yi}. A possible non-perturbative gauge-invariant composite operator is
\begin{align}
    O_{\mu}^{\bar{a}}(x) &=\tr\Bigg[\frac{\sigma^{\bar{a}}}{2}\,\frac{X(x)^\dagger}{\sqrt{\det X(x)}}\,D_\mu\,\frac{X(x)}{\sqrt{\det X(x)}}\Bigg]\;,
    \label{eq:opcont}
\end{align}
where $\bar{a}=1,2,3$, is an index belonging to the custodial symmetry group. In principle, such an operator requires genuine non-perturbative methods to evaluate\footnote{There is also the issue of operator mixing, to which we will turn in Section \ref{s:opchoice}.}. This will be done later using lattice methods to address our primary aim.

\subsection{The properties of the weak vector boson in GIPT}

For our secondary aim, GIPT \cite{Frohlich:1980gj,Frohlich:1981yi,Maas:2017wzi} allows for an analytical evaluation of this operator. This requires to switch to a gauge with explicit vacuum expectation value, e.g., a 't Hooft gauge \cite{Bohm:2001yx}. Then, we split the matrix $X$ as
\begin{align}
    X(x) &= v\,\Omega + \chi(x)\;,
    \label{eq:split}
\end{align}
where $\Omega\in\mathrm{SU}(2)$, $v$ is the vacuum expectation value (vev), and $\chi(x)$ is a matrix-representation of the fluctuations. The operator \pref{eq:opcont} is then expanded in both the vev $v$ and the couplings. Keeping for the moment all orders in the coupling, the leading order in $v$ yields \cite{Frohlich:1980gj,Maas:2017wzi}
\begin{align}
    O_{\mu}^{\bar{a}}(x) &= -\mathrm{i}\,g\,\tr\Big[\frac{\sigma^{\bar{a}}}{2}\Omega^\dagger\frac{\sigma^a}{2}\Omega \Big]\,W_\mu^a(x) + \mathcal{O}\big(v^{-1}\big) \equiv C^{\bar{a}a}\,W_\mu^a(x) + \mathcal{O}(v^{-1})\;,
    \label{eq:fmsops}
\end{align}
with $C^{\bar{a}a}=-\mathrm{i}\,g\,\tr\Big[\frac{\sigma^{\bar{a}}}{2}\Omega^\dagger\frac{\sigma^a}{2}\Omega \Big]$. If $\Omega=\mathbb{1}$, which is just a special gauge choice, then $C^{\bar{a}a}=-\mathrm{i}\,g\,\delta^{\bar{a}a}$. This shows that to leading order in $v$ the composite operator is equivalent to all orders in the couplings $g$ and $\lambda$ to the elementary gauge boson operator, thereby mapping custodial indices to gauge indices.

For the propagator in position-space this yields
\begin{align}
    \big\langle O_\mu^{\bar{a}}(x)\,O_\nu^{\bar{b}}(y)\big\rangle &= C^{\bar{a}a}\,C^{\bar{b}b}\,\big\langle W_\mu^a(x)\,W_\nu^b(y)\big\rangle\phantom{_\mathrm{tl}} + \mathcal{O}\big(v^{-1}\big)\label{eq:fmspropt} \\
    &= C^{\bar{a}a}\,C^{\bar{b}b}\,\big\langle W_\mu^a(x)\,W_\nu^b(y)\big\rangle_\mathrm{tl} + \mathcal{O}\big(v^{-1},g,\lambda\big) \;,\label{eq:fmsprop}
\end{align}
where in the first line \pref{eq:fmspropt} still the full propagator appears. Thus, at leading order in $v$ and to all orders in $g$ and $\lambda$ the poles on the left-hand side and the right-hand side coincide, and on-shell properties of both particles match.

In the second line \pref{eq:fmsprop} at leading-order in the double expansion of the right-hand-side correlator yields that this pole is at $m_{\textrm{W}}$. Thus, the physical vector triplet has the same mass as the unphysical gauge triplet, which has been explicitly confirmed in lattice calculations \cite{Maas:2012tj,Maas:2013aia}. This can be repeated for the whole standard-model \cite{Frohlich:1980gj,Frohlich:1981yi,Maas:2017wzi}, yielding always a map between a physical (custodial) state to one of the elementary (gauge-dependent) particles with coinciding poles. So far, this has been confirmed on the lattice also for the Higgs \cite{Maas:2012tj,Maas:2013aia}.

On-shell, this result remains true to all orders in $g$ and $\lambda$ because of the leading term in \pref{eq:fmspropt}. However, the subleading orders in $v$ could affect this result, especially off-shell. Therefore, we will determine both sides of \pref{eq:fmspropt} in Section \ref{s:prop}, and compare them, as they only differ in the sub-leading terms of the $v$ expansion.

\subsection{Form factor} \label{s:vertexintro}

Not only the particles have been investigated experimentally in detail, but also their interactions \cite{pdg}. Our primary aim is therefore the computationally\footnote{While having a three-point interaction with the physical scalar would reduce the Lorentz structure it would introduce disconnected contributions, which would substantially increase the numerical cost.} simplest such interaction. This will be the 3-vector vertex of the physical vector bosons, defined in momentum space as
\begin{align}
    V_{\mu\nu\rho}^{\bar{a}\bar{b}\bar{c}}(p,q,k) &= \big\langle O_\mu^{\bar{a}}(p)\,O_\nu^{\bar{b}}(q)\,O_\rho^{\bar{c}}(k)\big\rangle\;,
    \label{eq:vertex}
\end{align}
\no where momentum conservation implies $p+q+k=0$. Choosing a different custodial basis, this would be the physical version of the $W^+W^-Z$ vertex. However, this form is more convenient for the lattice calculations, especially as all three particles are identical without QED.

A general 3-vector vertex such as \eqref{eq:vertex} can be decomposed in 14 Lorentz-tensor structures $T^{(i)}_{\mu\nu\rho}(p,q,k)$ and, for the custodial SU(2) symmetry, one custodial rank three tensor \cite{Ball:1980ax}, i.e.,
\begin{align}
    V_{\mu\nu\rho}^{\bar{a}\bar{b}\bar{c}}(p,q,k) &= \epsilon^{\bar{a}^\prime\bar{b}^\prime\bar{c}^\prime}\,D_{\mu\mu^\prime}^{\bar{a}\bar{a}^\prime}(p)\,D_{\nu\nu^\prime}^{\bar{b}\bar{b}^\prime}(q)\,D_{\rho\rho^\prime}^{\bar{c}\bar{c}^\prime}(k)\,\sum_{i=1}^{14}\Gamma^{(i)}(p,q,k)\,T_{\mu^\prime\nu^\prime\rho^\prime}^{(i)}(p,q,k)\;,
    \label{eq:vertexdecomp}
\end{align}
where $D^{\bar{a}\bar{b}}_{\mu\nu}(p)$ is the Fourier transformation of the position-space propagator \eqref{eq:fmsprop}, and the $\Gamma^{(i)}(p,q,k)$ are the form factors or dressing functions. We will here concentrate on a single one of it, which will be motivated by our secondary aim, GIPT.

Performing the same steps as in Section \ref{s:propintro} for \pref{eq:vertex} yields
\begin{align}
    V_{\mu\nu\rho}^{\bar{a}\bar{b}\bar{c}}(p,q,k) &= C^{\bar{a}a}\,C^{\bar{b}b}\,C^{\bar{c}c}\,\big\langle W_\mu^a(p)\,W_\nu^b(q)\,W_\rho^c(k)\big\rangle\phantom{_\mathrm{tl}} + \mathcal{O}\big(v^{-1}\big)\label{eq:fmsvertext}\\
    &= C^{\bar{a}a}\,C^{\bar{b}b}\,C^{\bar{c}c}\,\big\langle W_\mu^a(p)\,W_\nu^b(q)\,W_\rho^c(k)\big\rangle_\mathrm{tl} + \mathcal{O}\big(v^{-1},g,\lambda\big)\;.\label{eq:fmsvertex}
\end{align}
Thus, the gauge-invariant 3-vector vertex is to lowest order in $v$ identical to the tree-level 3-W vertex from standard perturbation theory. This especially implies that to leading order in $v$ the derived coupling constant will be identical, and no anomalous three-gauge-coupling (atgc) arises. This is consistent with experimental measurements \cite{pdg}. Again, this may change in subleading orders in $v$, and therefore we determine and compare both sides of Equation \pref{eq:fmsvertext} in Section \ref{s:ff} separately to assess the size of these.

At leading order in $v$, $g$, and $\lambda$ Equation \pref{eq:fmsvertex} implies furthermore that the gauge-invariant vertex should be given to this order by the tree-level 3-$W$ vertex, i.e., should have only the tensor structure
\begin{align}
    T_{\mu\nu\rho}^{(1)} &= (q-k)_\mu\,\delta_{\nu\rho}+(k-p)_\nu\,\delta_{\mu\rho}+(p-q)_\rho\,\delta_{\mu\nu}\;.
\end{align}
We will therefore concentrate here entirely on this one. This will keep the technical complications for the lattice calculations at a handable level.

We therefore consider here not \pref{eq:vertex}, but rather project to the tree-level basis element and solve for the tree-level form factor. This yields
\begin{align}
\begin{split}
    \Gamma(p,q,k)&=\Gamma^{(1)}(p,q,k) \\
    &= \frac{\big\langle \Gamma_{\mu\nu\rho}^{\bar{a}\bar{b}\bar{c}}(p,q,k)\,O_\mu^{\bar{a}}(p)\,O_\nu^{\bar{b}}(q)\,O_\rho^{\bar{c}}(k)\big\rangle}{\Gamma_{\mu\nu\rho}^{\bar{a}\bar{b}\bar{c}}(p,q,k)\,D_{\mu\mu^\prime}^{\bar{a}\bar{a}^\prime}(p)\,D_{\nu\nu^\prime}^{\bar{b}\bar{b}^\prime}(q)\,D_{\rho\rho^\prime}^{\bar{c}\bar{c}^\prime}(k)\,\Gamma_{\mu^\prime\nu^\prime\rho^\prime}^{\bar{a}^\prime\bar{b}^\prime\bar{c}^\prime}(p,q,k)}\;,
\end{split}
\label{eq:ratio}
\end{align}
where we used the definition $\Gamma_{\mu\nu\rho}^{\bar{a}\bar{b}\bar{c}}(p,q,k)=\epsilon^{\bar{a}\bar{b}\bar{c}}\,T_{\mu\nu\rho}^{(1)}(p,q,k)$. 
Using \eqref{eq:fmspropt} and \eqref{eq:fmsvertext} shows that \eqref{eq:ratio} is to leading order in $v$ just the amputated tree-level 3-W vertex. At leading order in $v$, $g$, and $\lambda$, \pref{eq:fmsprop} and \pref{eq:fmsvertex} imply $\Gamma=1$. Thus, any deviation from $\Gamma=1$ measures the deviation from double leading-order GIPT.

\subsection{Radius}

From form factors radii of particles can be defined \cite{williams1991nuclear,Pacetti:2015iqa}. Consider a spherical particle with charge density $\rho(r)$, where $r$ is the radial coordinate. The form factor is given by
\begin{align}
    \Gamma(p) &= \frac{\int \mathrm{d}^3r\,\rho(r)\,\mathrm{e}^{\mathrm{i}\,{\bf p}\cdot{\bf r}}}{\int\mathrm{d}^3r\,\rho(r)} = \frac{1}{p}\frac{\int\limits_0^\infty\mathrm{d}\,r\,\rho(r)\,\sin\big(p\,r\big)}{\int\limits_0^\infty \mathrm{d}r\,r^2\rho(r)} = 1 - \frac{p^2}{6}\big\langle r^2\big\rangle + \cdots\;,
    \label{eq:taylorform}
\end{align}
where we used a Taylor expansion of the $\sin$-function in the third equality, and $p$ is a characteristic scale of the momentum configuration. We defined $\big\langle r^2\big\rangle = \int\limits_0^\infty \mathrm{d}r\,r^2\rho(r)\,r^2/\int\limits_0^\infty \mathrm{d}r\,r^2\rho(r)$, which is the expectation value of the square of the radius of the considered particle.

Therefore, the square of the weak radius is given by
\begin{align}
    \big\langle r^2\big\rangle &= -6\,\left.\frac{\mathrm{d}\Gamma(p)}{\mathrm{d}p^2}\right|_{p^2=0}\;.
    \label{eq:radius}
\end{align}
The definition \pref{eq:taylorform} used actually the three-momentum to motivate the definition. The final formula \pref{eq:radius} then defines the radius at space-like, i.e., Euclidean, four-momenta \cite{Dissertori:2003pj}, which we will employ. Note, that if the right-hand side should be negative this formally implies an imaginary radius, which is at odds with any particle-like interpretation\footnote{Note that a negative radius can occur if the probed particle is uncharged as, e.g., happens for the electric radius of the neutron. But here all probed particles are always charged, and thus a positive radius is expected.}. E.g., gluons in Yang-Mills theory have such an imaginary radius \cite{Maas:2011se}.

However, because of the finite volume zero momentum is not accessible. Thus, \pref{eq:radius} cannot be determined directly. We attempted to determine the radius using a fit to a Taylor expansion of $\Gamma$, but found that the results were strongly affected by finite-volume artifacts. On the other hand, for weak interactions of particles with a single resonance with mass $m_W^2$ it is expected that the form factor, normalized to one at $p^2=0$, behaves like \cite{Pacetti:2015iqa}
\be
\Gamma(p^2,p^2,p^2)=\frac{1}{N}\left(a\frac{m_W^2}{p^2+m_W^2}+b\right)\label{vmd}
\ee
\no where $b$ is added to include the possibility to have just a tree-level behavior, and $N=N(a,b,m_W^2)$ is the normalization ensuring $\Gamma(0)=1$. This has a pole at time-like momenta at the resonance. The quantity $a/N$ parameterized the coupling strength to this pole in this channel, from which we will define a reduced coupling strength $\alpha=(a/N)^2/(4\pi)$. This expression fits the results quite well, and we will use this analytical form to determine the radius \pref{eq:radius} analytically, as well as $\alpha$, in Section \ref{ss:radius}. The results from fitting using a Taylor expansion were consistent, though much less reliable, with the fit ansatz \pref{vmd}.

\subsection{Subtleties} \label{s:opchoice}

There are two subtleties, which need to be addressed.

The first is with respect to tensor structures. In the Landau gauge limit the gauge fields satisfy $\pd^\mu W_\mu^a=0$. Thus, the propagators are transverse \cite{Bohm:2001yx} and only four out of the 14 tensor structures in \pref{eq:vertexdecomp} have non-zero dressing functions \cite{Ball:1980ax}. However, for gauge-invariant quantities this can necessarily not have any implications. Especially, a massive propagator can have, and will be seen below has, a longitudinal, poleless part \cite{Itzykson:1980rh}, and can have non-vanishing form-factors for all 14 tensor structures in the decomposition \pref{eq:vertexdecomp}.

This seems to kill at first sight already statements like \pref{eq:fmsprop} and \pref{eq:fmsvertex}, as they would be violated in Landau gauge, and it seems subleading terms in \pref{eq:fmspropt} and  \pref{eq:fmsvertext} would be necessary. However, this is not true. Statements that correlation functions have tensor indices are only an mnemonic, rather than literally true. Any quantity not invariant under Lorentz symmetry actually vanishes, as no direction is preferred. Rather, only fully contracted quantities, like \pref{eq:ratio}, can be non-zero, even for the propagators. Thus, agreement or disagreement has to be considered independently for each form factor. Here, we consider two form factors for the propagator, the transverse one and the longitudinal one, and the form factor \pref{eq:ratio} for the vertex. And indeed, the entirely gauge-dependent longitudinal part of the propagator will disagree at leading order, as the physical one is non-zero and the gauge-dependent one in the here chosen Landau gauge is zero. This statement is true even for the expression \pref{eq:fmspropt}, and thus at leading order in $v$ and all orders in $g$ and $\lambda$. The longitudinal information of the physical operator requires thus in this gauge sub-leading orders in $v$.

The second is connected with the choice \pref{eq:opcont}. Of course, any operator with the same quantum numbers will mix with this one. This has been studied for this channel on the lattice \cite{Maas:2014pba,Wurtz:2013ova}. Fortunately, most operators in this channel appear to decay very quickly to the ground-state \cite{Maas:2014pba}, and thus describe very well asymptotically a single physical vector particle. Thus, for the sake of having well-defined asymptotic states to really speak of testing a radius, such a choice is sufficient, and the remaining effects are contained in the finite-volume systematics for the lattice formulation used here.

Conversely, this implies that there exists an operator basis for every channel in which the ground-state is contained only in a single operator, the perfect ground-state operator ${\tilde O}$. In Euclidean space-time, the operator will generate a propagator
\be
D(p)=\big\langle \tilde{O}_i(p)^\dagger\,\tilde{O}_i(p)\big\rangle=\frac{Z}{p^2+m^2}\;,
\label{eq:physprop}
\ee
where $i$ is a suitable multiindex, $m$ is the groundstate mass in this channel, and $Z$ is a wave-function renormalization. This implies that if all propagators in the channel are renormalized as $D(0)=1/m^2$, this propagator is a lower bound at all momenta for all physical correlators \cite{Seiler:1982pw}. This will be relevant below in Section \ref{s:prop} and the analysis of the relation \pref{eq:fmspropt}.

\section{Technical details}\label{s:lattice}

The lattice simulations use the same setup as in \cite{Maas:2012tj,Maas:2013aia,Maas:2014pba}, to which we refer for technical details. For the gauge-fixed calculations we use a Landau gauge, in which we average over all possible orientations of the vev, making use of the fact that for the gauge bosons no direction is preferred. See \cite{Maas:2013aia} for details and how this relates to gauges with an explicit direction for the vev. This also allows us to average all quantities over custodial and gauge indices, effectively increasing our statistics. Especially, in this gauge \pref{eq:fmsvertext} and \pref{eq:fmsvertex} hold to very good accuracy on-shell \cite{Maas:2012tj,Maas:2013aia}.

\begin{table}[t!]
\centering
{\small
\begin{tabular}{c|ccccccc}
 \hline
 \hline

 Name & $\beta$ & $\kappa$ & $\lambda$ & $a^{-1}$ [GeV] & $m_{1^-_3}$ [GeV] & $m_{0^+_1}$ & $\alpha(200\text{ GeV})$ \cr
 \hline
 A1 & 2.7984 & 0.2927 & 1.317 & 453 & 80 & 124 & 0.605 \cr
 A2 & 2.7987 & 0.2953 & 1.267 & 335 & 80 & 122 & 0.506 \cr

 A3 & 2.3634 & 0.3223 & 1.066 & 151 & 80 & 131 & 0.558 \cr
 \hline
 B1 & 4.2000 & 0.2736 & 1.000 & 438 & 80 & 129 & 0.106 \cr
 B2 & 4.0000 & 0.3000 & 1.000 & 255 & 80 & 118 & 0.211 \cr 
 \hline
 \hline
\end{tabular}}
\caption{\label{tab:para}The parameters of the lattice setup. The masses of the physical vector bosons and the physical scalar are obtained as in \cite{Maas:2014pba}, and the running coupling $\alpha$ in the miniMOM scheme \cite{vonSmekal:2009ae} as in \cite{Maas:2013aia}. The lattice spacing $a^{-1}$ is set such that the physical vector boson has a mass of 80.375 GeV. The statistical errors are suppressed, but are at the sub-percent level for all quantities except $m_{0_1^+}$, for which it is at the order of a few percent.}
\end{table}

The lattice parameters we use are listed in Table \ref{tab:para}. As there are three bare parameters three physical quantities are needed to characterize them, which we choose to be the physical scalar singlet mass, the physical vector triplet mass, and the running weak gauge coupling in the miniMOM scheme \cite{vonSmekal:2009ae} at 200 GeV. These have been determined using the methods in \cite{Maas:2014pba,Maas:2013aia}. We fix the two masses such that they roughly correspond to the masses in the standard model, knowing that because of the absence of QED in our simulations they can be at best adequate to the level of $\sim$10 GeV. The lattice spacing is set by requiring that the vector triplet has a mass of 80.375 GeV, and thus this mass is identical for all parameter sets in Table \ref{tab:para}.

The sets are grouped into a set with rather large gauge coupling, much larger than in the standard model, to enhance subleading effects (set A), and one with smaller coupling to see how the effects change when moving towards more standard-model-like couplings (set B). In both cases multiple lattice spacings have been used. The individual sets are most likely not part of a line-of-constant physics, but are the best approximations to these right now we have available from the existing phase diagram scans in \cite{Maas:2014pba,Maas:unpublished}. At any rate, it will be seen that the results are rather comparable within the sets, and for some of the results even among both sets. Thus, this seems to be a minor issue. Similar hints of independence of the actual choice of the line-of-constant physics have also been seen in other spectroscopically channels in \cite{Maas:2014pba}. For the present exploratory study this appears adequate, though a systematic investigation along several lines of constant physics remains desirable for quantitatively reliable statements.

\begin{table}[t!]
\centering
\begin{tabular}{cccccc}
 \hline
 \hline
 
 Set & A1 & A2 & A3 & B1 & B2 \cr
 \hline
 $8^4$ & 1032200 & 1032200 & 1032200 & 1032200 & 1032200 \cr
 $12^4$ & 680625 & 687500 & 687500 & 687500 & 687500 \cr
 $16^4$ & 254700 & 254700 & 254700 & 254700 & 254700 \cr
 $20^4$ & 198400 & 198400 & 198400 & 197408 & 198400 \cr
 \hline
 \hline
\end{tabular}
\caption{\label{tab:c1}Number of configurations for physical quantities.}
\end{table}

\begin{table}[t!]
\centering
\begin{tabular}{cccccc}
\hline
 \hline
 
 Set & A1 & A2 & A3 & B1 & B2 \cr
 \hline
 $8^4$ & 85462 & 142128 & 147500 & 147500 & 147500 \cr
 $12^4$ & 119885 & 137500 & 137500 & 137500 & 137500 \cr
 $16^4$ & 62186 & 63683 & 63700 & 63700 & 63700 \cr
 $20^4$ & 60878 & 65092 & 66200 & 65869 & 66200 \cr
 \hline
 \hline
\end{tabular}
\caption{\label{tab:c2}Number of gauge-fixed configurations for gauge-dependent quantities.}
\end{table}

For every set of parameter the volumes $8^4$, $12^4$, $16^4$, and $20^4$ have been simulated. As it turned out, very large amounts of statistics, listed in Tables \ref{tab:c1} and \ref{tab:c2}, are necessary, which prevented us from going to even larger volumes in this exploratory study. However, for most of our results the statistical error currently still dominates in comparison to systematic errors, as will be seen below.

The final ingredient is the choice of operators. For the gauge-dependent $W$ propagator and 3-$W$ vertex we employ the same methods as in \cite{Maas:2013aia}, to which we refer for details. For the gauge-invariant operator \pref{eq:opcont}, from which we will build the gauge-invariant propagator and 3-vector vertex, we choose
\begin{align}
    O_{\mu}^{\bar{a}}(x) &= \tr\Bigg[\frac{\sigma^{\bar{a}}}{2}\,\frac{X(x)^\dagger}{\sqrt{\det X(x)}}\,U_\mu(x)\,\frac{X(x+\hat{\mu})}{\sqrt{\det X(x+\hat{\mu})}}\Bigg]\;,
    \label{eq:oplat}
\end{align}
where $U_\mu(x)=\mathrm{e}^{\mathrm{i}a W_\mu(x)}$ is the gauge-link. The operator \eqref{eq:oplat} is transformed into momentum space using standard lattice methods for both the propagator and the vertex \cite{Cucchieri:2006tf}. As it turns out, lattice improvements will be needed for both the propagator and the vertex, which will be discussed in Sections \ref{ss:latprop} and \ref{ss:latff}. Note, that because on the lattice the fields are rescaled \cite{Rothe:2005nw,Maas:2013aia}, for both the propagators and the vertices this rescaling has to be reversed to obtain continuum expressions. There are also factors of the coupling constants, which need to be removed to ensure $\Gamma=1$ in the tree-level case. This has been done.

\section{Propagator}\label{s:prop}

\subsection{Lattice aspects}\label{ss:latprop}

For the gauge-dependent $W$ propagator we refer for details of the implementation to \cite{Cucchieri:2006tf,Maas:2013aia}. In our choice of Landau gauge there is only a transverse dressing function left, which will be denoted $D_{\textrm{W}}$ in the following. For the gauge-invariant one, we decompose the propagator in a transverse part $D_{\textrm{T}}$ and a longitudinal part $D_{\textrm{L}}$ as
\begin{align}
    D_{\mu\nu}^{\bar{a}\bar{b}}(p) &= \delta^{\bar{a}\bar{b}}\,\Big(T_\mn(p)\, D_\mathrm{T}(p)\,+L_\mn(p)\, D_\mathrm{L}(p)\Big)
    \label{eq:giprop}\;.
\end{align}
As discussed in Section \ref{s:opchoice}, the dressing functions $D_{\textrm{T}}$ and $D_{\textrm{L}}$ need to be obtained by contraction with corresponding tensors from the correlation function $\big\langle O_{\mu}^{\bar{a}}(p)\,O_{\nu}^{\bar{b}}(p)^\dagger\big\rangle$, which is built from the lattice operator \pref{eq:oplat}. For this we use the lattice-improved projectors 
\begin{align}
    T_{\mu\nu}(p) &= \delta_{\mu\nu}-\frac{\sin\big(P_\mu\big)\sin\big(P_\nu\big)}{\sum\limits_{\rho=1}^4 \sin^2\big(P_\rho\big)} 
    \quad,\quad
    L_{\mu\nu}(p) = \frac{\sin\big(P_\mu\big)\sin\big(P_\nu\big)}{\sum\limits_{\rho=1}^4 \sin^2\big(P_\rho\big)} \;,
    \label{eq:projectors}
\end{align}
with $P_\mu=\pi n_\mu/L$, $n_\mu=0,1,\dots,N/2$, with $N$ being the extent of the lattice, and $p_\mu=2/a\sin(P_\mu)$ the improved continuum momentum. We will evaluate the physical propagator along a momentum with two non-vanishing components, i.e., along a plaquette diagonal, and with one non-vanishing momentum component, i.e., along an edge. These are the momentum configurations which will be needed in Section \ref{s:ff} for the projection \pref{eq:ratio}. For the gauge-dependent $W$ propagator several additional momentum configurations are used, as described in \cite{Cucchieri:2006tf}, reaching to much larger momenta.

\begin{figure}[!htb]
    \hspace*{-1.4cm}
    \includegraphics[width=1.2\textwidth]{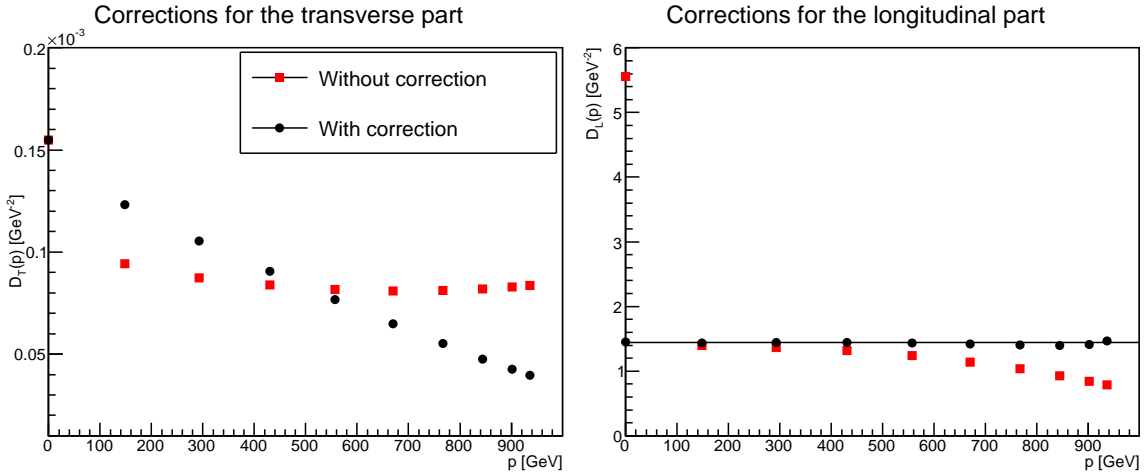}
    \caption{On the left-hand side we show the transverse part of the propagator and on the right-hand side the longitudinal part of the propagator as a function of momentum with two non-vanishing components. Results are plotted with (black circles) and without (red squares) lattice corrections for the largest lattice volume $20^4$ and for set A1.}
    \label{fig:propscorr}
\end{figure}

Unfortunately, it turns out that the physical propagator is much stronger affected by lattice artifacts than the $W$ propagator. This is shown in Figure \ref{fig:propscorr}. Especially noteworthy is that the longitudinal propagator is not constant, which is forbidden in the continuum limit \cite{Itzykson:1980rh}, and the transverse one raises at large momenta, which is equally forbidden for a physical correlation function \cite{Seiler:1982pw}. Both properties are thus clearly lattice artifacts. In fact, the expected behaviors are 

\begin{align}
    D_\mathrm{T}(p) &= \sum_i\frac{Z_i}{p^2 + m_{i}^2}\stackrel{p\to\infty}{\longrightarrow}\frac{Z}{p^2}
    \quad,\quad
    D_\mathrm{L}(p)=\text{const}\;.\label{eq:asymp}
\end{align}
We use a purely phenomenological approach to reduce these artifacts. To establish the correct asymptotic behavior we use
\begin{align}
    D_\mathrm{T}(p)\,p^2 &- A\,\sin(|P|)^\frac{3}{2}\;,
    \label{eq:latcorrtrans}
\end{align}
and adjust the parameter $A$ such that the result becomes constant for large momenta. For the longitudinal part we fit the lattice propagator by
\begin{align}
    B+C\,\cos(|P|)^{-\frac{5}{4}}\;.
    \label{eq:latcorrlong}
\end{align}
Then, only the second term of the fit is subtracted from $D_\mathrm{L}(p)$, and we set $D_\mathrm{L}(0) = B+C$. Furthermore, at zero momentum the decomposition in transverse and longitudinal part is not unique, and thus only the sum can be determined. Thus, at zero momentum also the so obtained $D_\mathrm{L}(0)$ has to be subtracted from $D_\mathrm{T}(0)$, which is in Figure \ref{fig:propscorr} not visible, as the propagators are renormalized to the value $1/(80.375\text{ GeV})^2$ at zero momentum.

The results of these corrections are also shown in Figure \ref{fig:propscorr}. The transverse part is now a monotonically decreasing function of momentum and the longitudinal part is approximately constant. Both corrections do, by construction, not affect the infrared behavior, as it should be for a pure lattice artifact\footnote{That this may look differently for the transverse propagator in Figure \ref{fig:propscorr} is because both have been renormalized.}. Of course, as already mentioned, this approach is just a phenomenological one and proper improvements could be obtained using lattice perturbation theory \cite{Rothe:2005nw}. However, given the involved six-point functions this appears overkill for the present exploratory study.

\subsection{Results}\label{ss:prop}

In the following we concentrate on the transverse propagator only, as the gauge-dependent one vanishes by construction and the longitudinal physical one is constant. We renormalize the transverse propagators by requiring $D_\mathrm{T}(0)=1/m_\mathrm{phys}^2$, with $m_\mathrm{phys}=80.375$ GeV, i.e., the experimentally determined mass of the physical vector boson \cite{pdg}. As both propagators turn out to be only weakly affected by the volume, we can concentrate here on the largest volume only.

\begin{figure}[!t]
    \hspace*{-1.4cm}
    \includegraphics[width=1.2\textwidth]{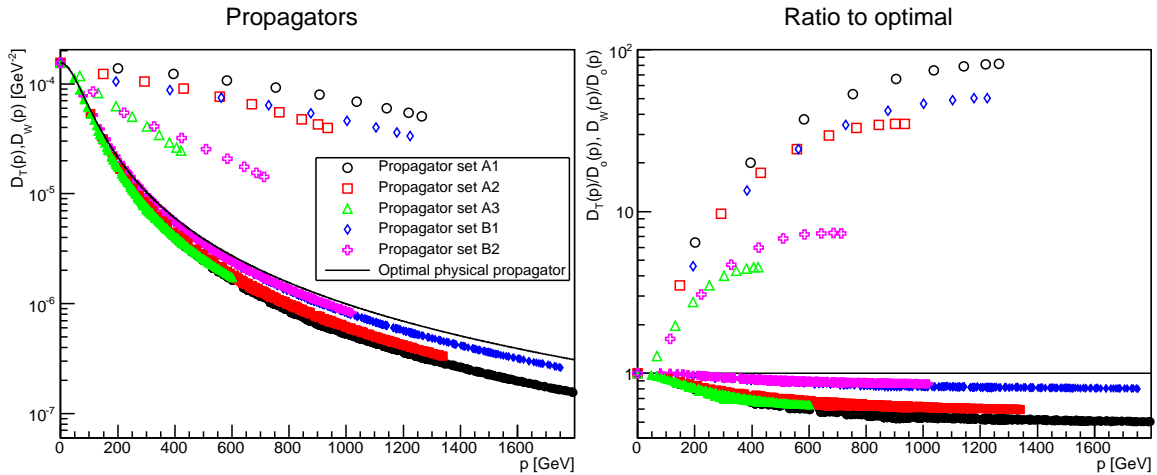}
    \caption{The transverse propagators as function of momentum for the largest lattice volume $20^4$ for all parameter sets available. The lattice-corrected transverse gauge-invariant propagators are shown as open symbols, the propagators of elementary fields are displayed with the corresponding closed symbols, and the optimal ground state propagator \eqref{eq:physprop} is also shown. On the right-hand side the ratio of the gauge-invariant as well as gauge-variant propagators to the optimal propagator are shown. All error bars are much smaller than the symbol sizes.}
    \label{fig:props}
\end{figure}

The lattice-improved physical transverse propagators for all sets and for the largest lattice volume $20^4$ are shown in Figure \ref{fig:props}. All the propagators lie well above the line for the physical ground state propagator \eqref{eq:physprop}, and satisfy ultimately at large momenta the behavior \pref{eq:asymp}. The deviation at small momenta is because the operator \pref{eq:opcont} does not coincide with the optimal ground-state operator ${\tilde O}$, but apparently excited states contribute, as discussed in Section \ref{s:opchoice}. Therefore, the transverse propagator of the bound state is a superposition of all these states, i.e., 
\begin{align}
    D_\mathrm{T}(p)=\sum_{i}\frac{Z_i}{p^2 + m_{i}^2}\;,
    \label{eq:superposprop}
\end{align}
where the $Z_i$ are the renormalized overlaps, and $m_i$ are the masses of the states contributing to this quantum number channel. It is also visible that the sets with finer lattice spacing pick up more massive modes in the sum \pref{eq:superposprop}, thus the later agreement with the asymptotic behavior \pref{eq:asymp}. As the two next levels due to scattering states are already of order 200 GeV and 240 GeV \cite{Maas:2014pba,Wurtz:2013ova} this is not unexpected given the lattice spacings listed in Table \ref{tab:para}. Other than these higher-state contaminations, the propagators show the expected physical behavior.

Note that in position-space the correlator decays over a much shorter time, i.e., about 3 lattice spacings, to the ground-state one than the mo\-men\-tum-space correlator suggests \cite{Maas:2014pba}. A different choice of basis, e.g., the one obtained from a variational analysis as in \cite{Maas:2014pba}, would yield better agreement with the ground state in momentum space, though would not improve the position-space properties substantially.

The situation is very different for the gauge-dependent $W$-propagator, also shown in Figure \ref{fig:props}. Here the behavior is grouped not with the lattice spacing, but with the gauge coupling. The smaller the gauge coupling, the longer the propagator remains close to the physical, optimal ground-state one. At large momenta, it starts to deviate below the optimal one. As noted in Section \ref{s:opchoice}, this implies that no physical state can be described by it. This conclusion is supported by observations of non-positive contributions to its corresponding spectral function \cite{Maas:2013aia} as well as the Oehme-Zimmermann superconvergence relation \cite{Oehme:1979ai}. However, as this is a gauge-dependent quantity, this is neither surprising nor problematic.

Concerning the primary objective it just shows that off-shell mixing becomes important. Given the observation in Section \ref{ss:radius} that the vector boson is relatively large, it is not surprising that a local operator like \pref{eq:opcont} picks up a lot of short-distance fluctuations. Using smeared operators, like in \cite{Maas:2014pba,Wurtz:2013ova}, it is straightforward to construct extended operators which are much better in agreement with the optimal ground-state one.

This leaves the following observation for the secondary objective, and thus of \pref{eq:fmspropt} and \pref{eq:fmsprop}. First of all, the tree-level expression in \pref{eq:fmsprop} coincides with the optimal one. The full $W$-propagator corresponds to the leading term in the vev expansion \pref{eq:fmspropt}, which includes all orders in $g$ and $\lambda$. As is seen in Figure \ref{fig:props}, this becomes a worse approximation the larger $g$ is. On the other hand, it s a very good approximation up to a few hundred GeV at small $g$. Thus, GIPT is well suited as an approximation of the ideal ground state propagator at sufficiently small $g$.

But this still shows that GIPT does not a very good job when literally considering the approximation \pref{eq:fmspropt}, because these are not statements about the optimal propagator, but about the ones given by the correlator of the operator \pref{eq:opcont}. Thus, the leading order in the $v$-expansion is not sufficient. As in higher orders immediately the scattering states appear at order $g^0$ and $\lambda^0$ \cite{Maas:2017wzi}, this would probably yield a much better agreement, than all orders in $g$ and $\lambda$, as was done here.

\section{Form-factor}\label{s:ff}

\subsection{Lattice aspects}\label{ss:latff}

For the 3-$W$ vertex the same techniques as in \cite{Maas:2013aia,Cucchieri:2006tf} have been used. Especially, in \cite{Maas:2013aia} the same vertex has been investigated, albeit at much smaller statistics. Out of the momenta configurations in \cite{Maas:2013aia} we will investigate here two particular ones. One is the symmetric one, in which the three momenta satisfy $p^2=q^2=k^2$. The other is a back-to-back configuration, in which one of the three momenta vanishes, e.g., $p=0$, which implies $k=-q$.

For the 3-vector vertex we use the same techniques\footnote{Note that this can be substantially improved to reduce lattice artifacts by having explicit separation of probe time and the times at which the probed particle is prepared \cite{Gockeler:2003ay,Koponen:2015tkr}. In the current exploratory study we will not do so, and consider these effects as a contribution to the finite-volume effects.} as in \cite{Cucchieri:2006tf}, except that we replace the $W$ field by the operator \pref{eq:oplat}, and the $W$ propagators by the vector triplet propagator \pref{eq:giprop}\footnote{ We have also tested what happens if we only use the transverse part of the vector triplet propagator, and found no significant quantitative deviations. This can be found explicitly in \cite{Raubitzek:2018graz}.}. In doing so, motivated by \pref{eq:fmsvertex}, we used in the ratio \pref{eq:ratio} the lattice version of the tree-level tensor structure of the 3-$W$ vertex \cite{Rothe:2005nw}, as is done for the 3-$W$ vertex itself \cite{Cucchieri:2006tf}, to remove the leading lattice artifacts. At the same time we did not apply the improvements \pref{eq:latcorrtrans} and \pref{eq:latcorrlong} to the involved physical propagators, as the ones appearing implicitly in the non-amputated nominator are not directly accessible to us.

\begin{figure}[!t]
    \hspace*{-1.4cm}
    \includegraphics[width=1.2\textwidth]{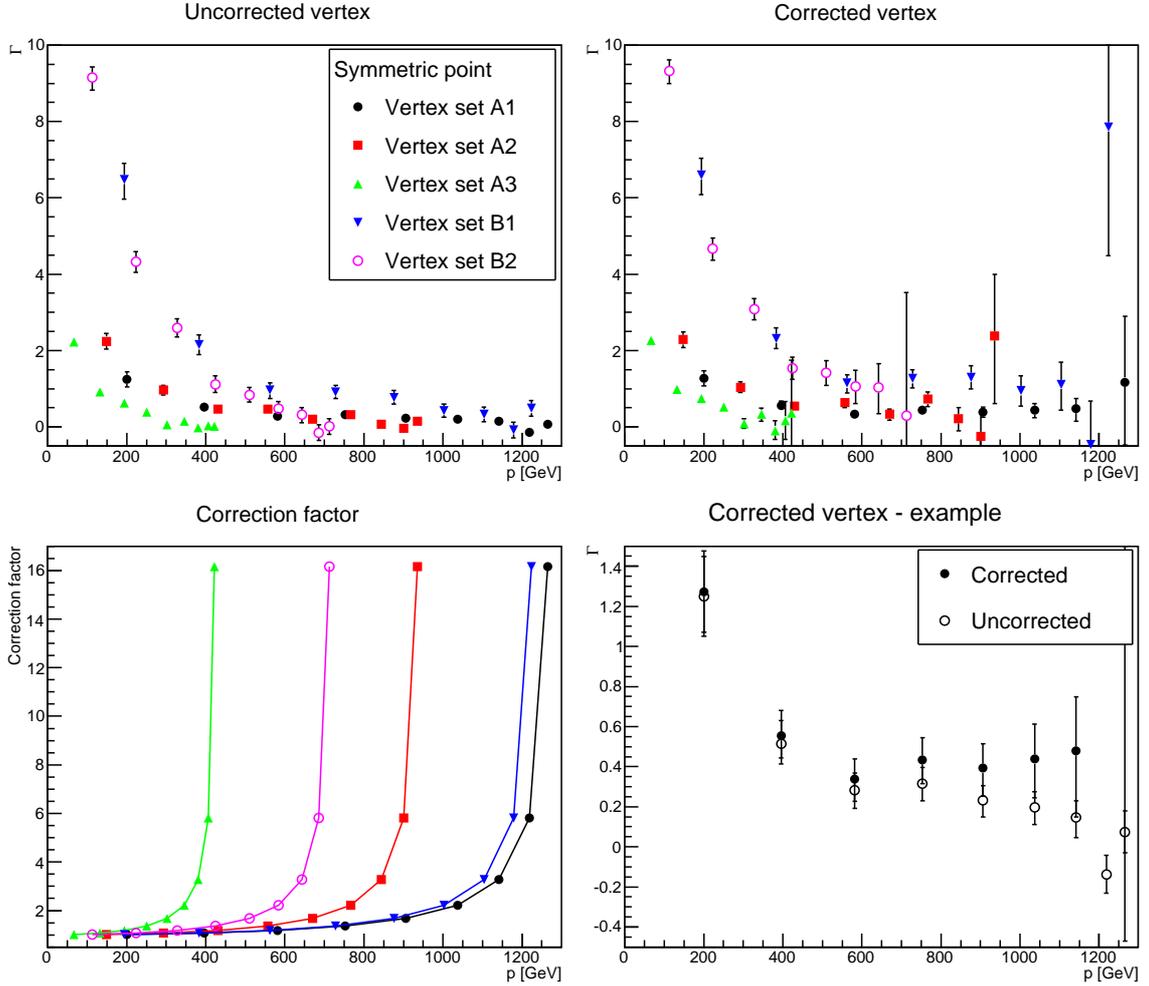}
    \caption{The form factor in the symmetric momentum configuration on the $20^4$ lattices for all parameter sets without (top-left panel) and with (top-right panel) the phenomenological lattice improvement. The bottom-left panel shows the correction factor alone, and the bottom-right panel a magnification of the corrected and uncorrected vertex for the A1 set.}
    \label{fig:vertexcorr}
\end{figure}

While this improvement program is sufficient for the 3-$W$ vertex it is not for the 3-vector vertex, as was already the case for the vector propagator in Section \ref{ss:latprop}. As is visible in Figure \ref{fig:vertexcorr}, the form factors drop towards zero at the largest respective physical momenta, i.e., at the same lattice momenta, irrespective of the lattice spacing. As the behavior is therefore clearly lattice-driven, this is an artifact. We correct this by the phenomenological correction
\begin{align}
\Gamma(p^2,q^2,k^2)\to \Gamma(p^2,q^2,k^2)\,\cos\left(|P|\right)^{-\frac{1}{2}}\cos\left(|Q|\right)^{-\frac{1}{2}}\cos\left(|K|\right)^{-\frac{1}{2}}\;,
\end{align}
the result of which is also shown in Figure \ref{fig:vertexcorr}. This yields an essentially constant behavior at large momenta, within statistical errors. Note that this does not change the infrared behavior, and is therefore not affecting the results for the radius in Section \ref{ss:radius}. Again, a better motivated improvement could be obtained from lattice perturbation theory \cite{Rothe:2005nw}, which in the present case involves a 9-point function, and is thus even more daunting than the 6-point case of the propagators.

\begin{figure}[!t]
    \hspace*{-1.4cm}
    \includegraphics[width=1.2\textwidth]{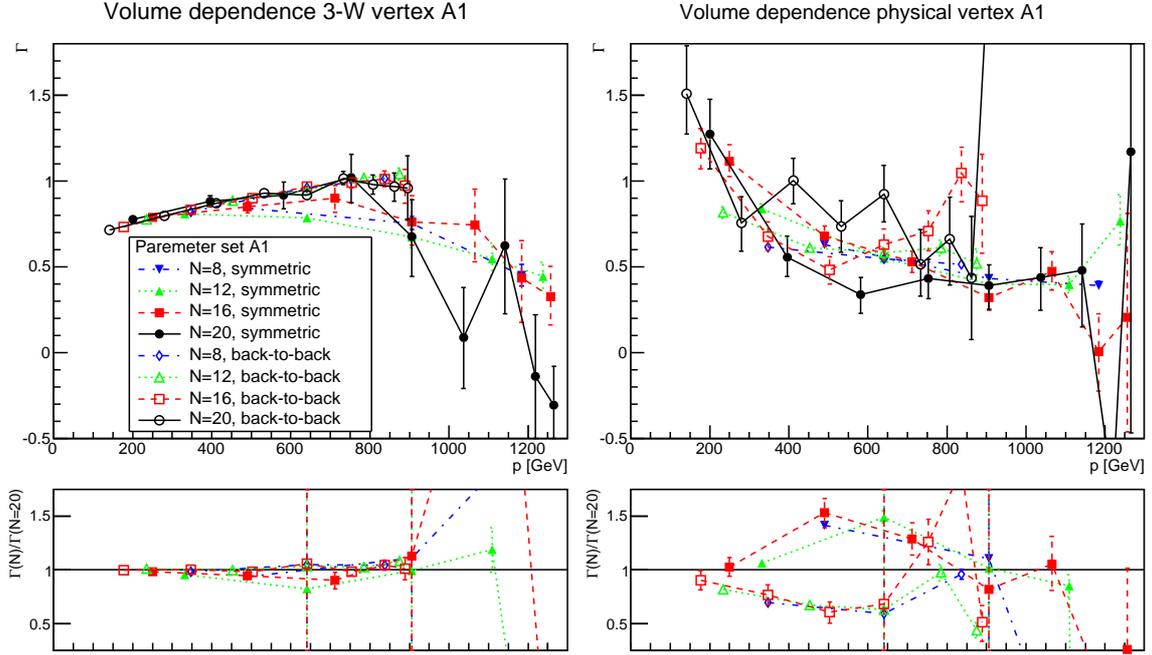}
    \caption{The 3-$W$ vertex form factor (left-hand side) and the physical 3-vector vertex form factor \pref{eq:ratio} (right-hand side) for set A1 for all volumes and momentum configurations. Here and hereafter, for the ratio the results from the largest volume have been linearly interpolated and the error determined by error propagation.}
    \label{fig:vertexa1}
\end{figure}

\begin{figure}[!t]
    \hspace*{-1.4cm}
    \includegraphics[width=1.2\textwidth]{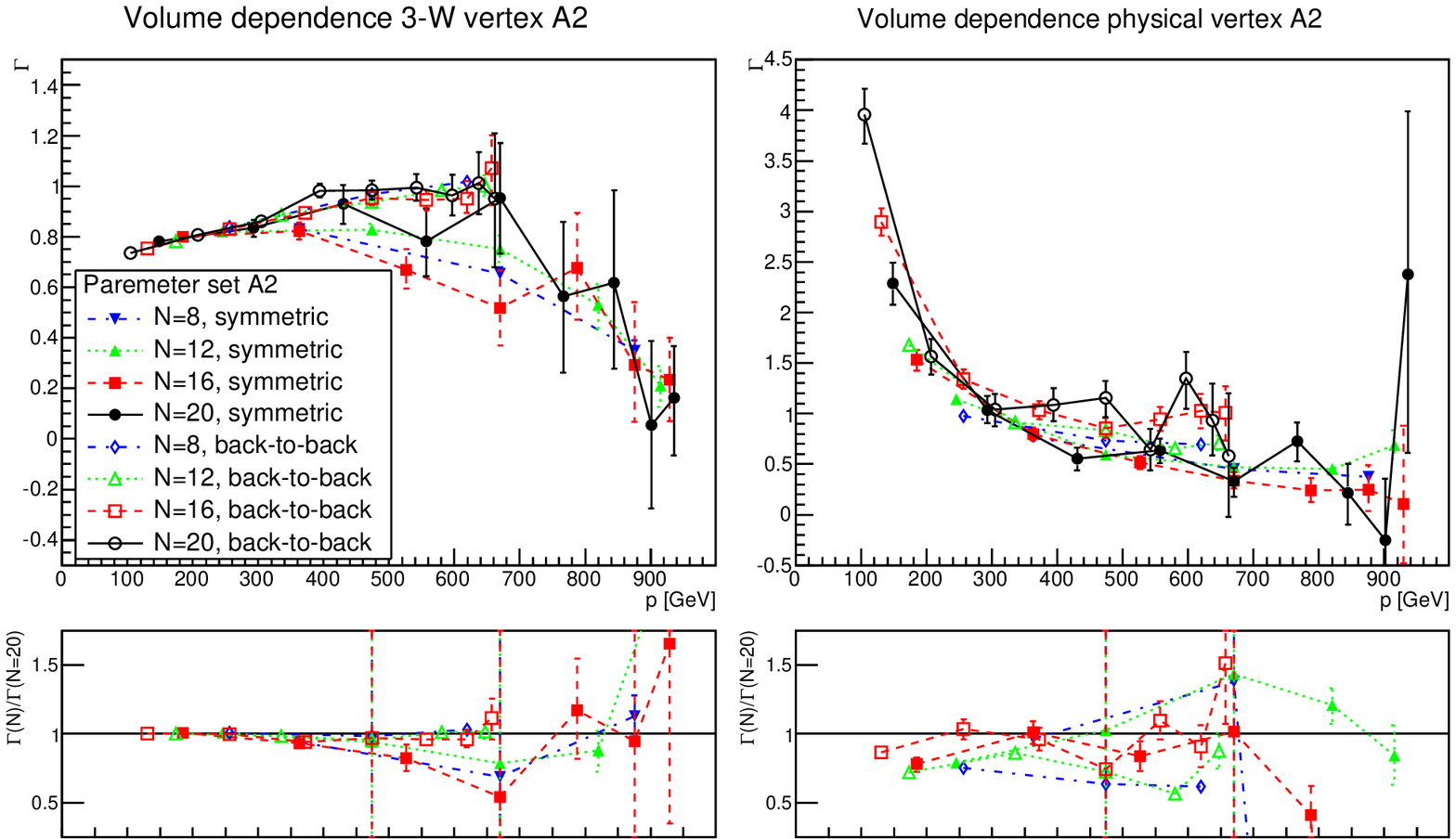}
    \caption{The 3-$W$ vertex form factor (left-hand side) and the physical 3-vector vertex form factor \pref{eq:ratio} (right-hand side) for set A2 for all volumes and momentum configurations.}
    \label{fig:vertexa2}
\end{figure}

\begin{figure}[!t]
    \hspace*{-1.4cm}
    \includegraphics[width=1.2\textwidth]{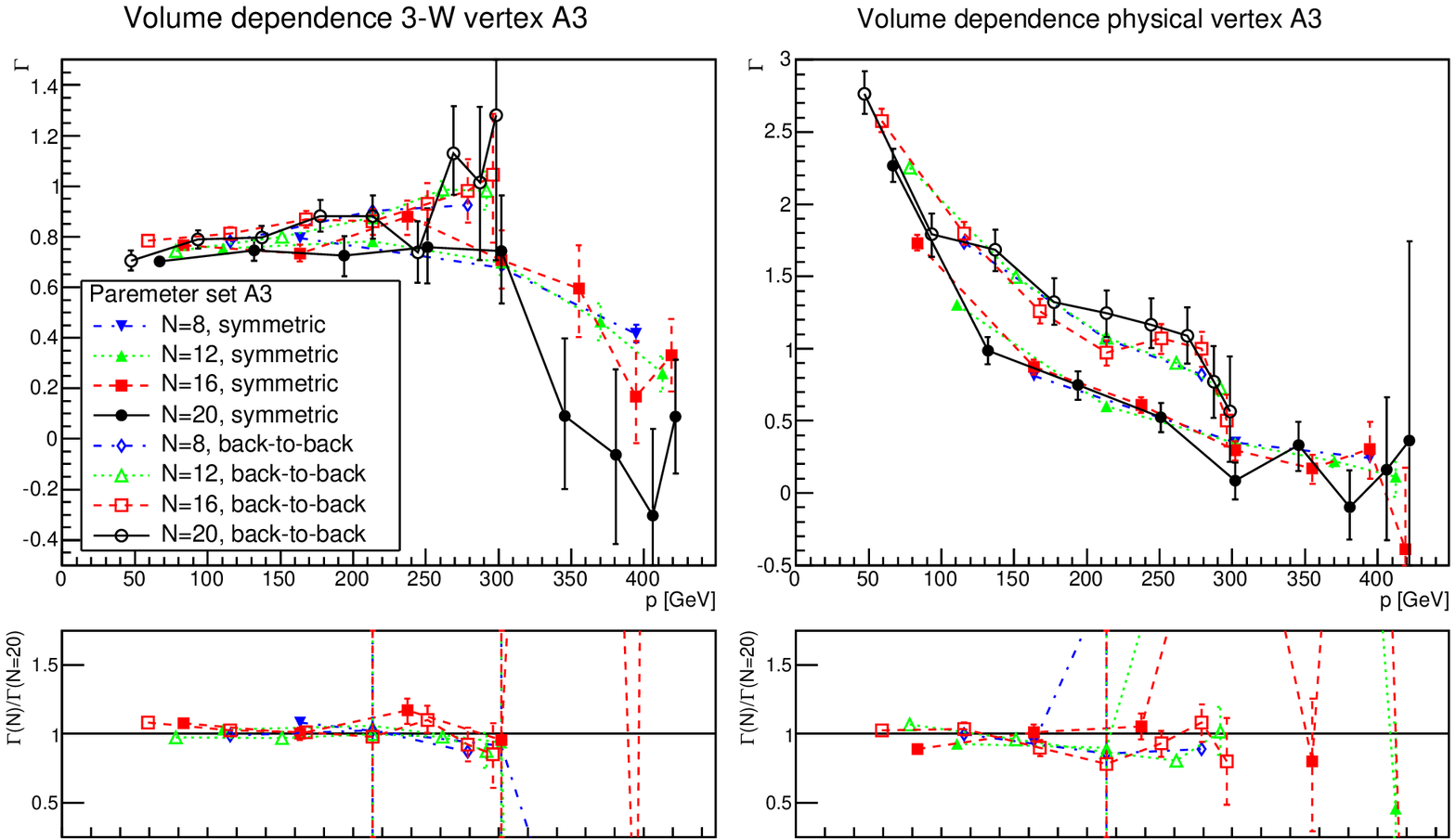}
    \caption{The 3-$W$ vertex form factor (left-hand side) and the physical 3-vector vertex form factor \pref{eq:ratio} (right-hand side) for set A3 for all volumes and momentum configurations.}
    \label{fig:vertexa3}
\end{figure}

\begin{figure}[!t]
    \hspace*{-1.4cm}
    \includegraphics[width=1.2\textwidth]{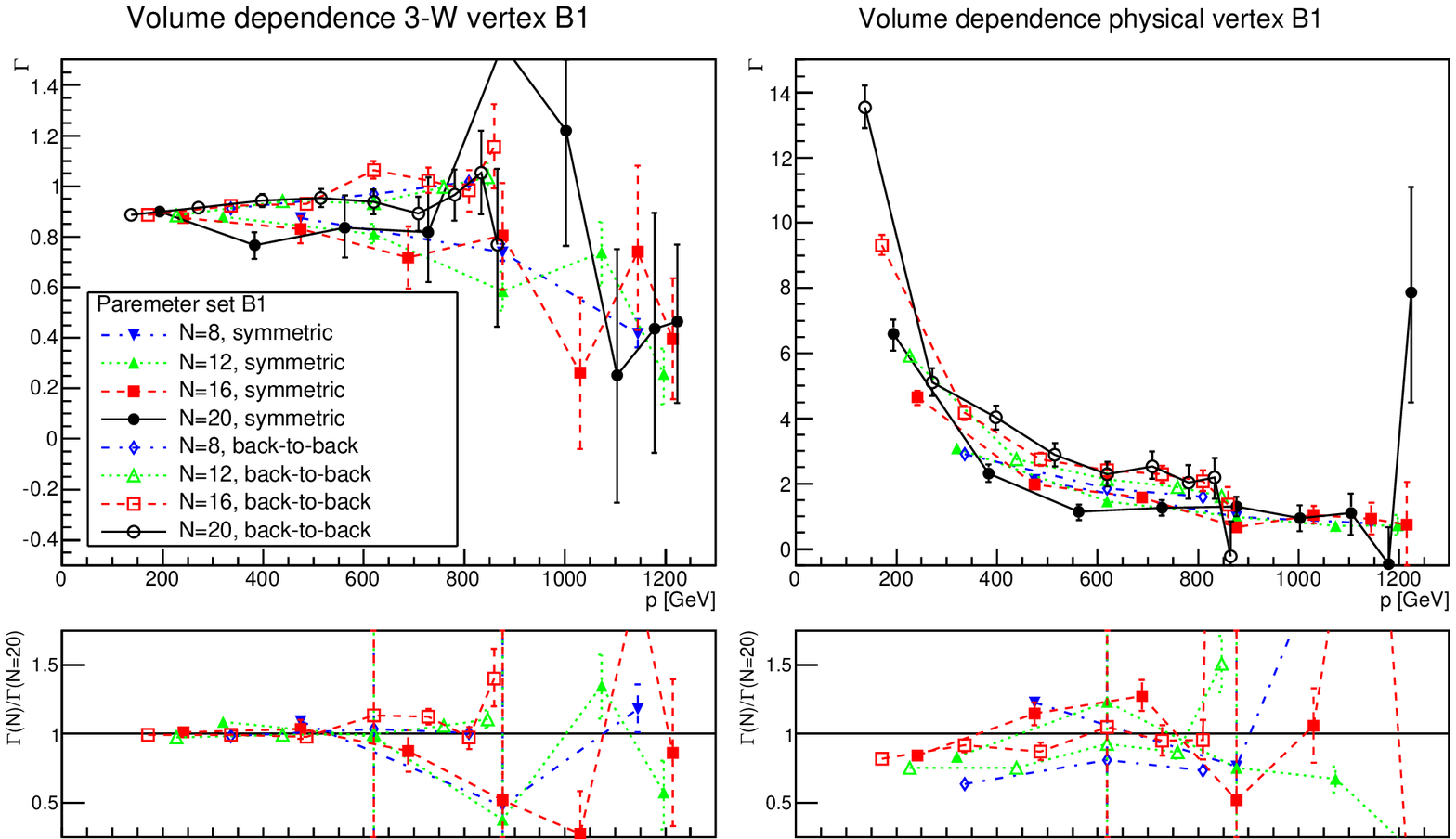}
    \caption{The 3-$W$ vertex form factor (left-hand side) and the physical 3-vector vertex form factor \pref{eq:ratio} (right-hand side) for set B1 for all volumes and momentum configurations.}
    \label{fig:vertexb1}
\end{figure}

\begin{figure}[!t]
    \hspace*{-1.4cm}
    \includegraphics[width=1.2\textwidth]{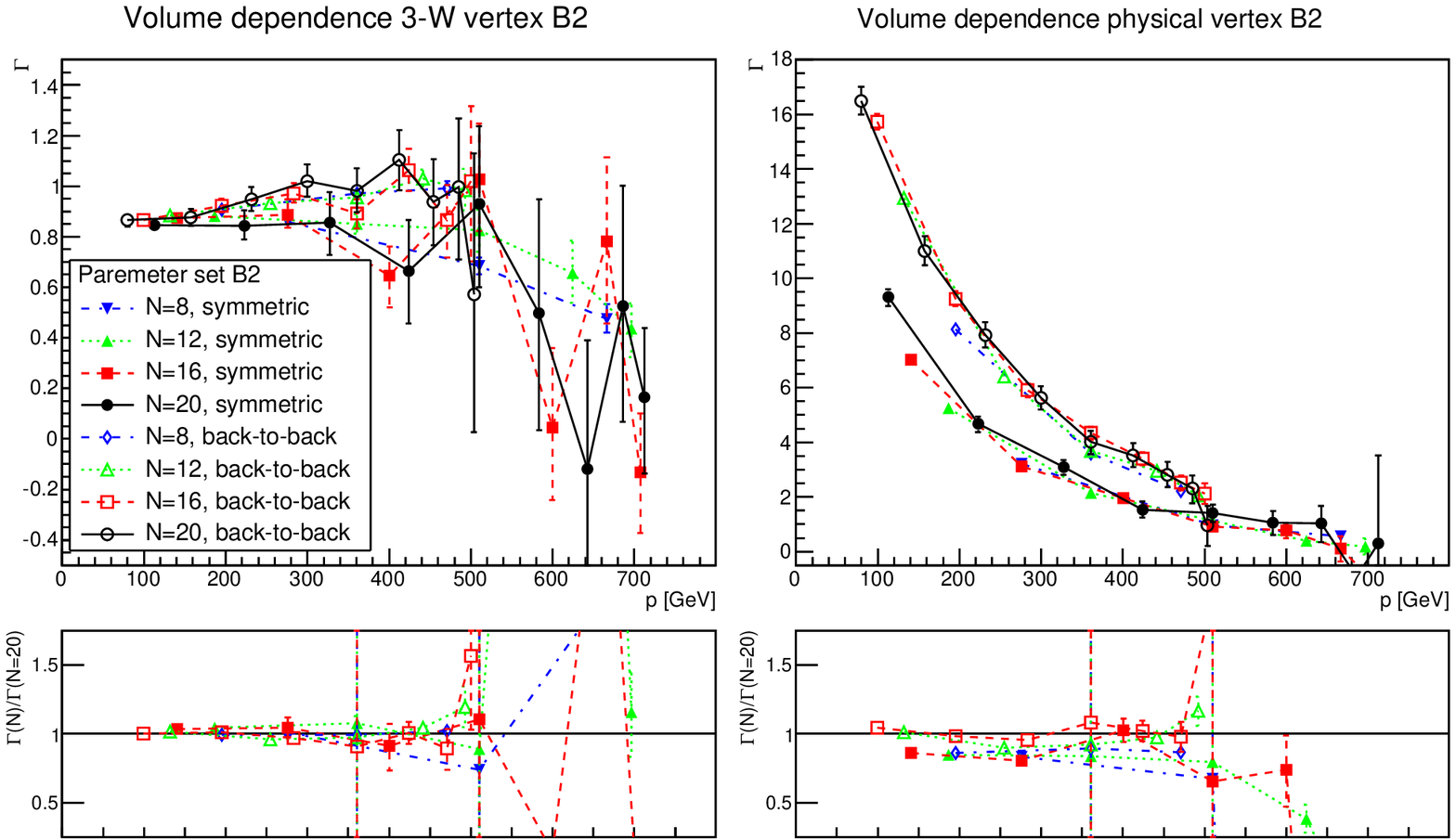}
    \caption{The 3-$W$ vertex form factor (left-hand side) and the physical 3-vector vertex form factor \pref{eq:ratio} (right-hand side) for set B2 for all volumes and momentum configurations.}
    \label{fig:vertexb2}
\end{figure}

To study the volume-dependence of both vertices we show the full set of results in Figures \ref{fig:vertexa1}-\ref{fig:vertexb2}. At large (lattice) momenta the form factors start to quickly drown in noise, which was observed for the gauge-dependent one already previously \cite{Cucchieri:2006tf,Maas:2013aia}. However, at medium and small momenta they show comparatively little volume dependence. Thus, in Section \ref{ss:ff} only the largest volume will be considered.

The different momentum configurations differ in some parts more than the statistical error, but show the same qualitative momentum-dependence. The vertices show on all sets the same qualitative behavior, but the 3-$W$ vertex and the physical vertex show pronounced different qualitative behavior, independent of the sets. The lattice spacing also shows no qualitative influence.

Note that the normalization in \pref{eq:ratio} vanishes when either all momenta vanish or any of them is the largest momentum on the lattice \cite{Cucchieri:2006tf}. Thus, these momenta are inaccessible.

\subsection{Form factor}\label{ss:ff}

\begin{figure}
    \includegraphics[width=0.5\textwidth]{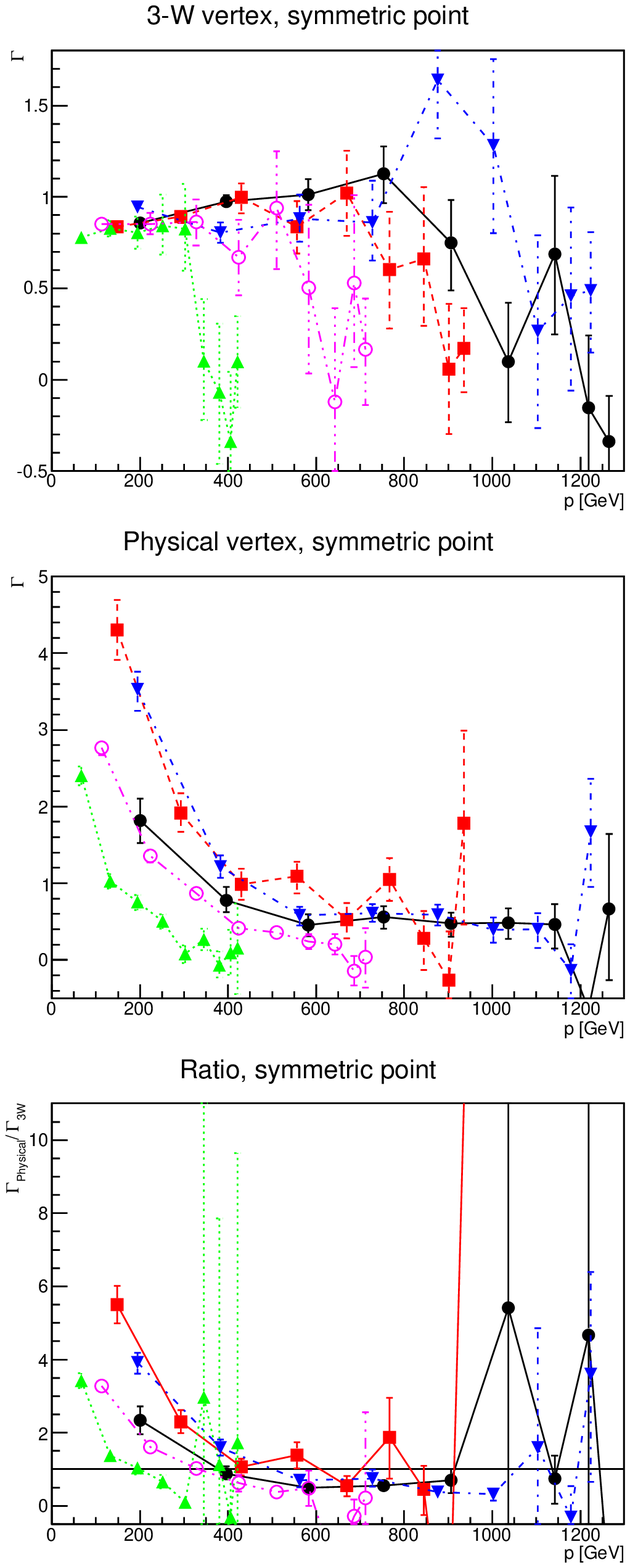}\includegraphics[width=0.5\textwidth]{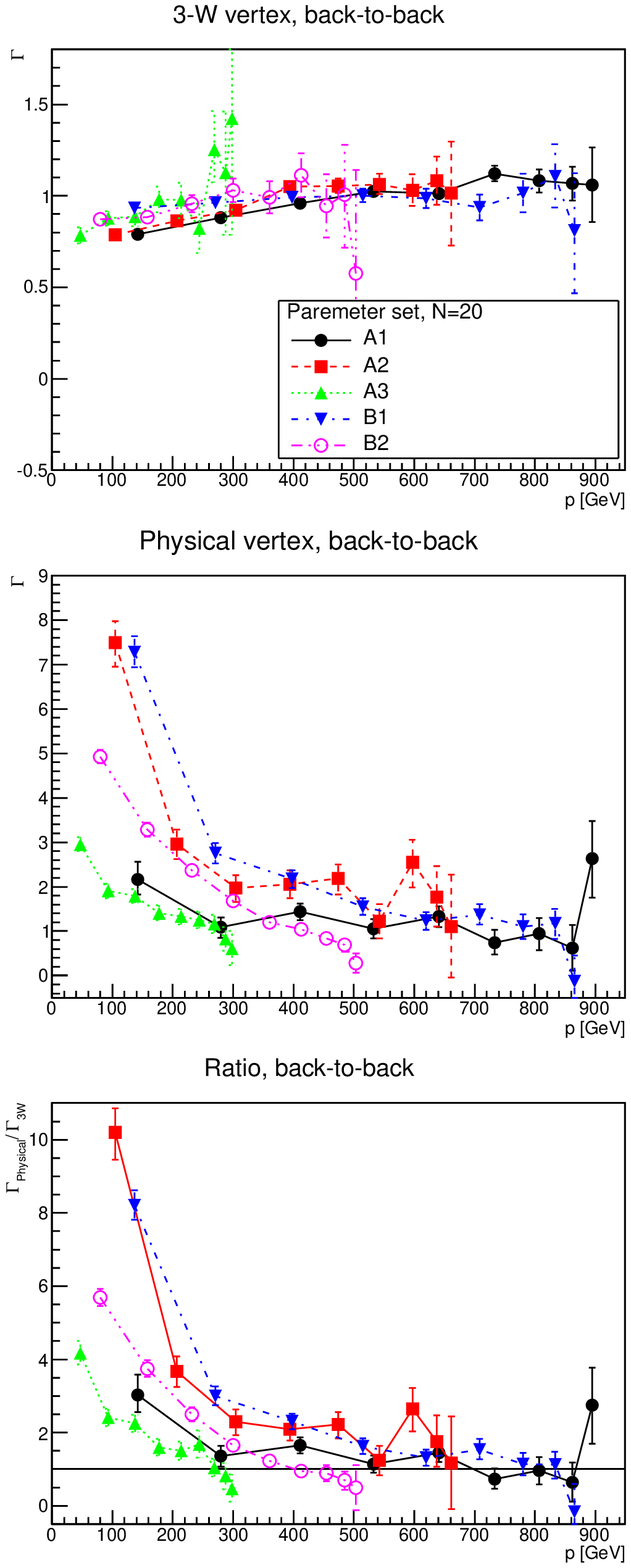}
    \caption{The 3-$W$ vertex form factor (top nales) and the physical 3-vector vertex form factor \pref{eq:ratio} (middle panels) in the symmetric (right panels) and back-to-back momentum configuration (left panels) for all sets on the largest volume. The bottom panels show the ratio of the physical form factor to the 3-$W$ form factor. The form factors have been renormalized in the back-to-back configurations to one at large momenta utilizing the fit results in Section \ref{ss:radius}.}
    \label{fig:vertex}
\end{figure}

The final results for the renormalized form factors in the symmetric and back-to-back momentum configurations are shown in Figure \ref{fig:vertex}. The qualitative behavior is the same for both momentum configurations, though they differ somewhat quantitatively.

First of all, the 3-$W$ vertex drowns quickly in noise at the largest (lattice) momenta. This also happens for the 3-vector vertex, but here the much larger statistics counteracts the effect somewhat. This effect is stronger for the symmetric momentum configuration, which extends to larger (lattice) momenta. Thus, the following will concentrate on the low and medium momentum behavior.

The results for the 3-$W$ vertex are essentially independent of the parameter sets. They show, as was already indicated at lower statistics in \cite{Maas:2013aia}, an almost tree-level form factor, which slowly decreases in the infrared. This is surprising as the running coupling for the set A is quite large, and larger deviations could have been expected.

The situation is somewhat different for the physical form factor. Here, the results group themselves into two sets for the stronger and weaker coupling. Both tend towards a constant at larger momenta, but the weaker interacting case deviates from this earlier towards the infrared. Still, both rise substantially in the infrared. As will be discussed in Section \ref{ss:radius} this can be attributed to the dominating pole at time-like momentum.

The ratio of both form factors shows that the essential constant dependency at large momenta is common to both, within the statistical errors. However, there is some angular dependence, manifesting itself in the different value of this constant ratio. At momenta below roughly 300-400 GeV, depending on the coupling strength, both form factors start to qualitatively deviate, with the physical rising while the 3-$W$ one slowly dropping. As discussed in Section \ref{s:vertexintro} this implies that only the physical 3-vector vertex shows a behavior which is consistent with the one expected for a physical particle, while the behavior of the 3-$W$ vertex is not. Again, this is neither surprising nor problematic, as the 3-$W$ vertex is gauge-dependent.

Concerning the primary objective, there is thus a qualitative and pronounced difference between the gauge-dependent 3-$W$ vertex and the physical 3-vector vertex at low momenta. Both are not equivalent observables.

Concerning the secondary objective, the quality of GIPT in \pref{eq:fmsvertext} and \pref{eq:fmsvertex}, this implies that the approximation in this case gets better at larger momenta, though it does not capture the angular dependence fully. Here it is not directly obvious if using the perfect operator ${\tilde O}$ rather than \pref{eq:oplat} would improve the situation, as the actually behavior for the optimal case is unknown. Nonetheless, this implies again that subleading orders in the $v$ expansion can become relevant in certain momentum regimes, this time at smaller momenta. The momentum regime where GIPT works becomes larger at smaller coupling, as it is seen that the deviation from a constant ratio occurs at higher momenta for the stronger coupled case.

Still, the behavior can be intuitively understood. At higher energies, the probe is more sensitive to the structure of the physical state than the physical state as a whole. At the same time, the structure of the physical probe itself becomes more relevant. As both probe and probed state have to leading order the $W$ boson as constituent, the interaction of 3 $W$s is probed at high energy, and thus the leading term in $v$ of GIPT, being just the 3-$W$-interaction, is indeed giving already a not so bad estimate. On the other hand, at low momentum the whole state is probed, which, as seen in the next section, is dominated by the close-by pole of the physical, stable vector particle at time-like momenta.

\subsection{Radius}\label{ss:radius}

\begin{figure}
    \includegraphics[width=0.5\textwidth]{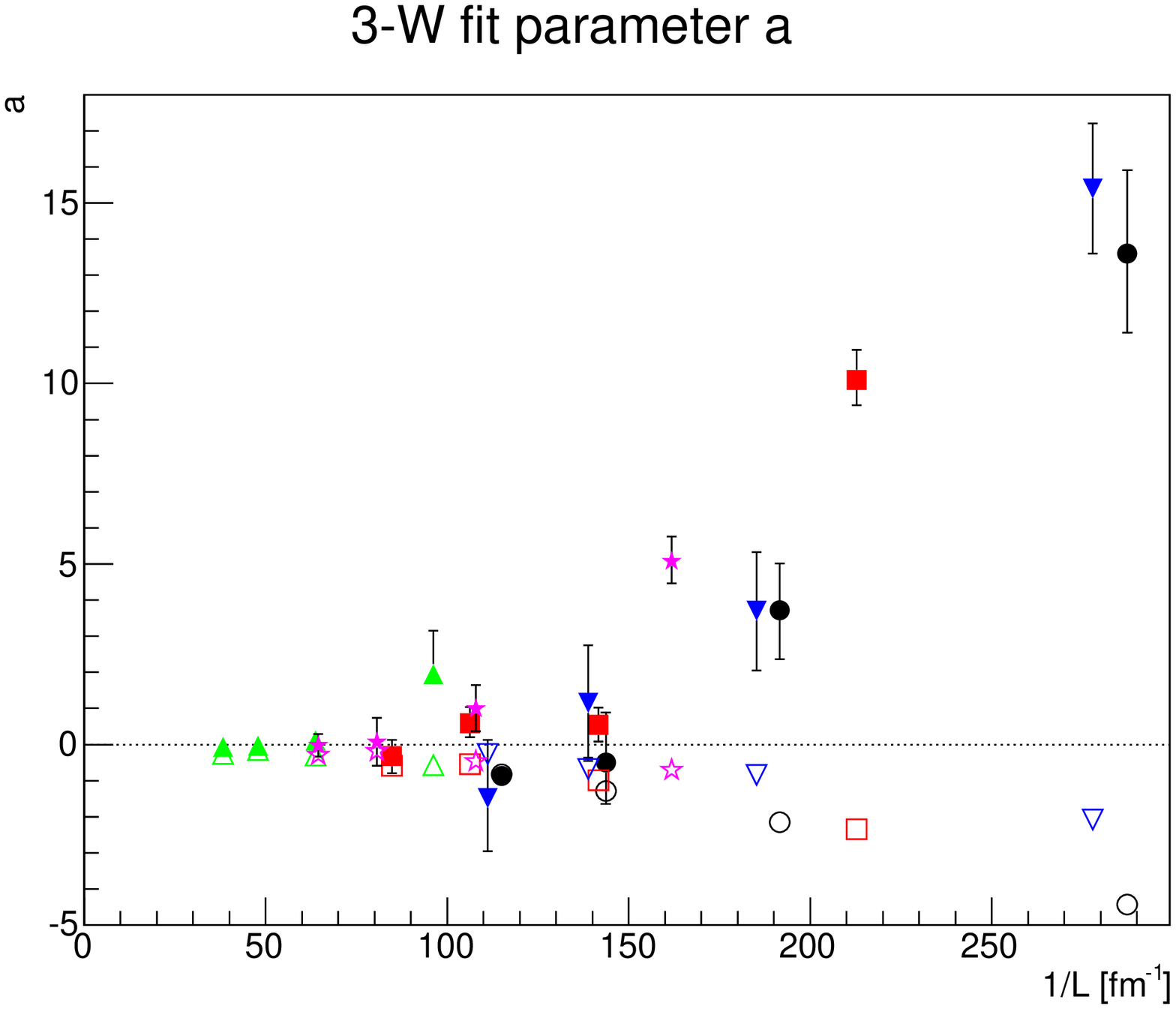}\includegraphics[width=0.5\textwidth]{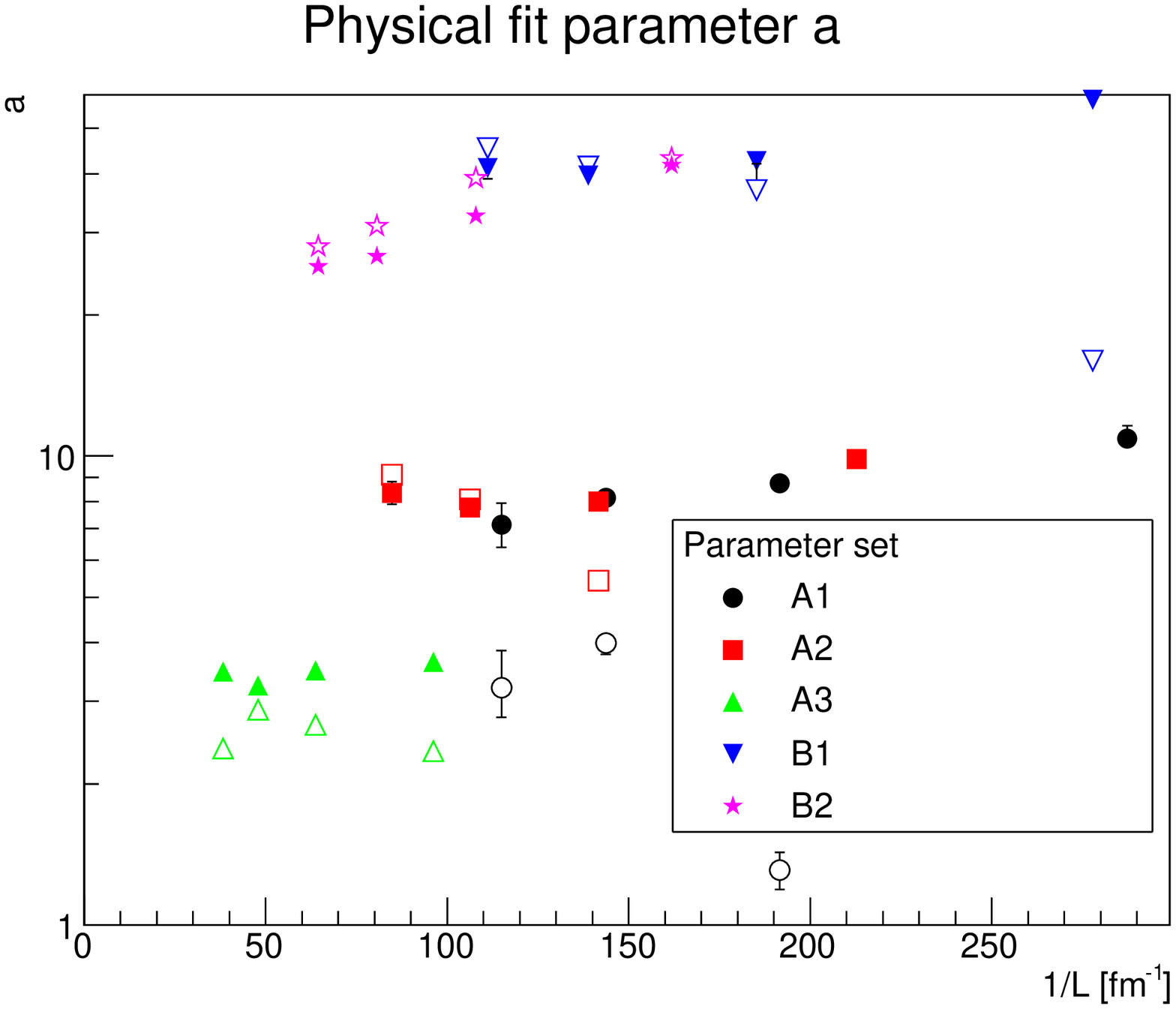}\\
    \includegraphics[width=0.5\textwidth]{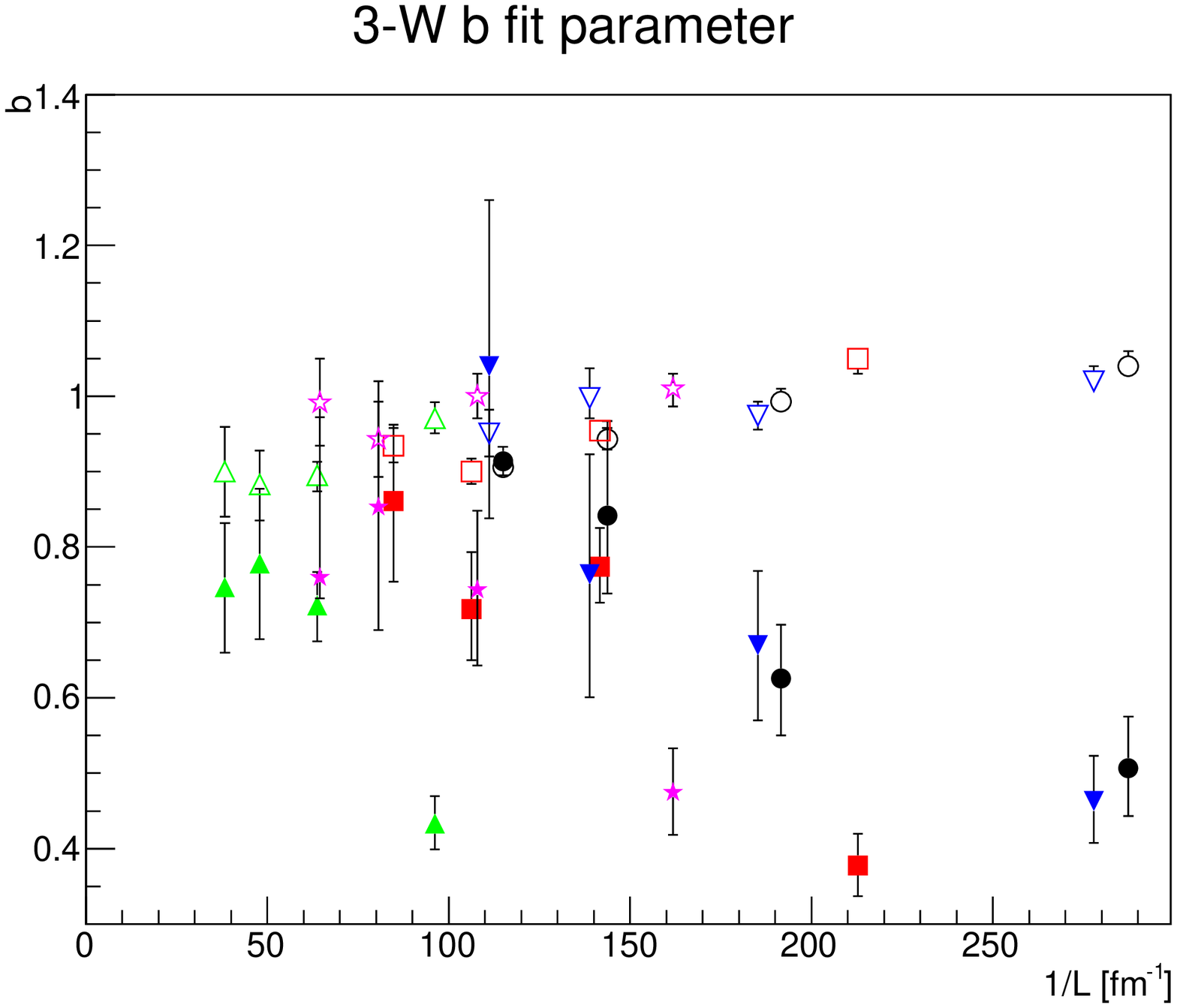}\includegraphics[width=0.5\textwidth]{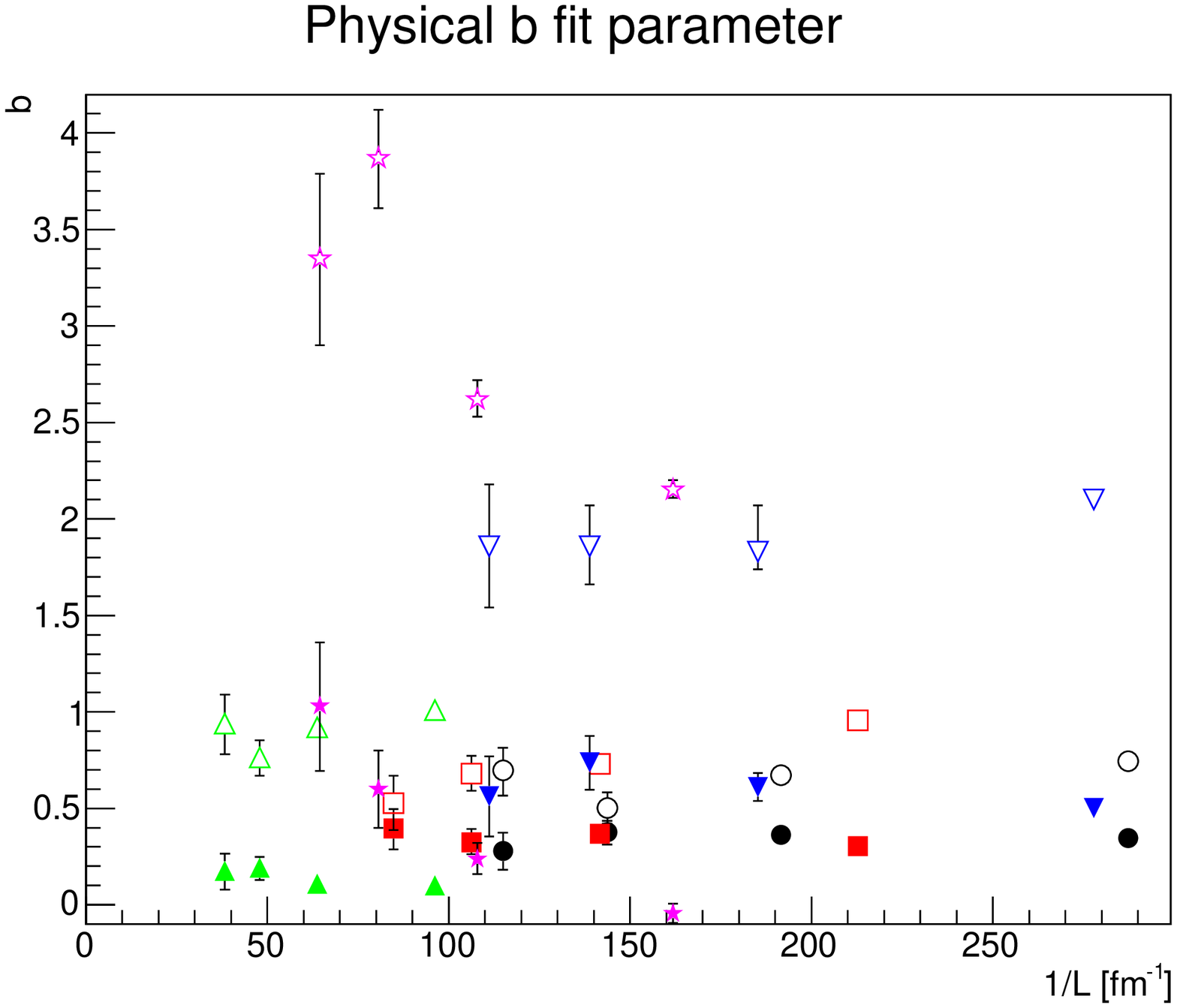}\\
    \caption{The fit parameters $a$ (top panels) and $b$ (bottom panels) for the 3-$W$ form factor (left panels) and the physical form factor (right panels) for all lattice setups. Open symbols are from the back-to-back configuration and closed symbols from the symmetric configuration.}
    \label{fig:fitab}
\end{figure}

\begin{figure}
    \includegraphics[width=\textwidth]{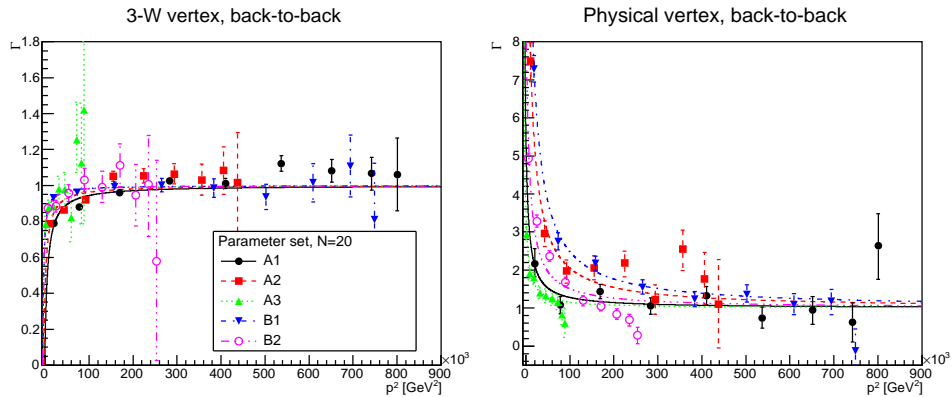}
    \caption{Comparison of the fit \pref{vmd} using the fit parameters shown in Figure \ref{fig:fitab} compared to the data in the back-to-back configuration for the 3-W form factor (left panel) and physical form factor (right panel). Note that the plot is against $p^2$ down to the pole located at $p^2=-m_W^2$ to illustrate its closeness.}
    \label{fig:fit}
\end{figure}

The resulting fit parameters $a$ and $b$ using the fit ansatz\footnote{For the back-to-back configuration the right-hand momentum in the fit was chosen to equal the non-zero momentum.} \pref{vmd} are shown in Figure \ref{fig:fitab}. Except for the set B2 all physical form factors have been fitted with a $\chi^2$/dof of at most 2 for the largest lattice volumes. The situation is not as good for the 3-$W$ form factor, as the deviation from tree-level requires large volumes to establish beyond statistical error. Thus, only the sets A3 and B2, with the respective largest physical volumes of the sets A and B, could be fitted with a $\chi^2$ below 2. A comparison of the fits to less noisy back-to-back data is shown in Figure \ref{fig:fit}, which nicely illustrates how the physical form factor is affected by the close-by pole at $-m_W^2$.

The fit parameters for the 3-$W$ vertex show substantial finite-volume effects for small volumes, but eventually converge to rather similar values. Especially, they ultimately yield an $a<0$, and thus a negative residue at the pole. This is once more a behavior expected for an unphysical particle. Conversely, the parameters for the physical vertex show less volume-dependence, but a much wider spread across the different sets.

\begin{figure}
    \includegraphics[width=0.5\textwidth]{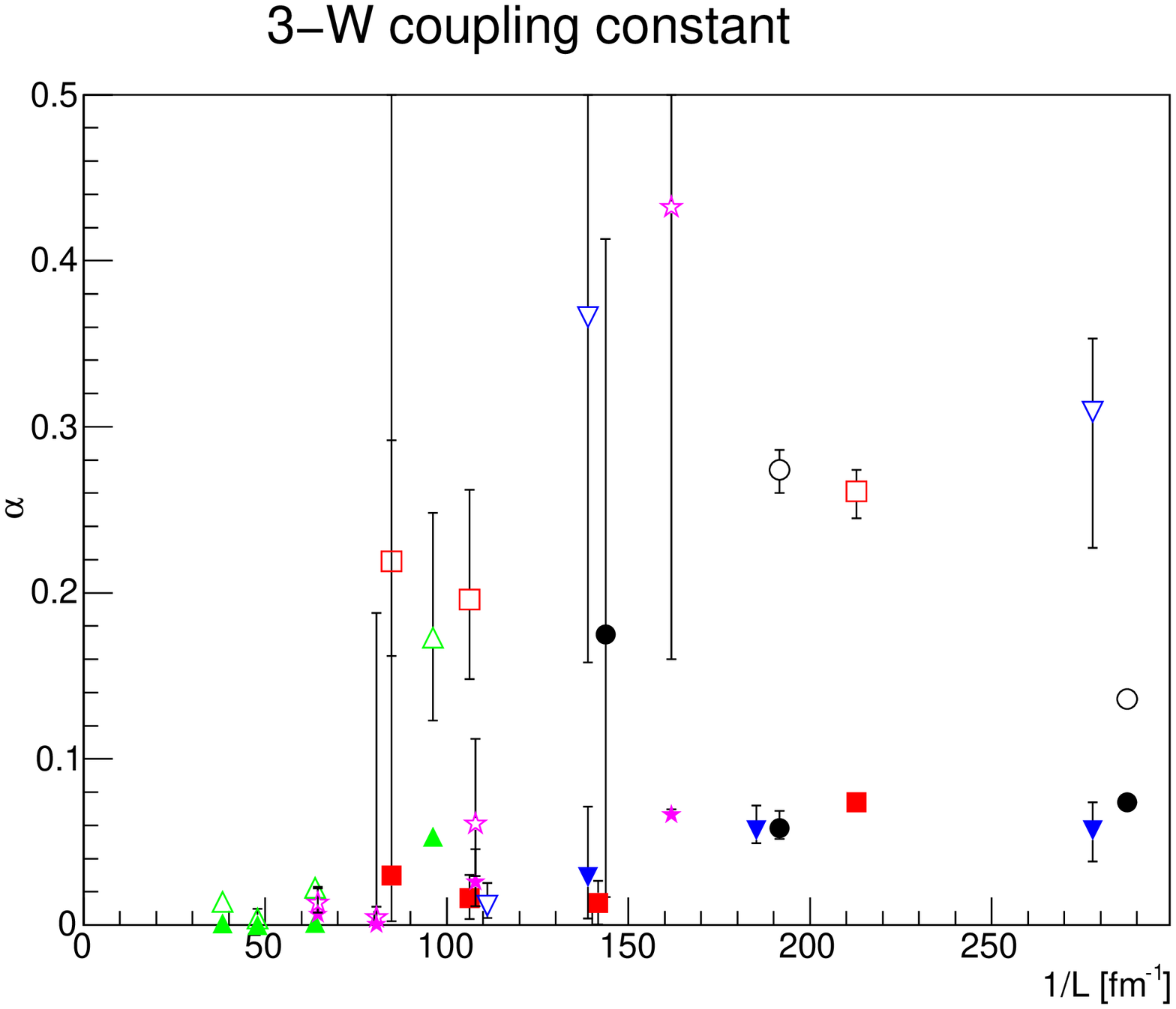}\includegraphics[width=0.5\textwidth]{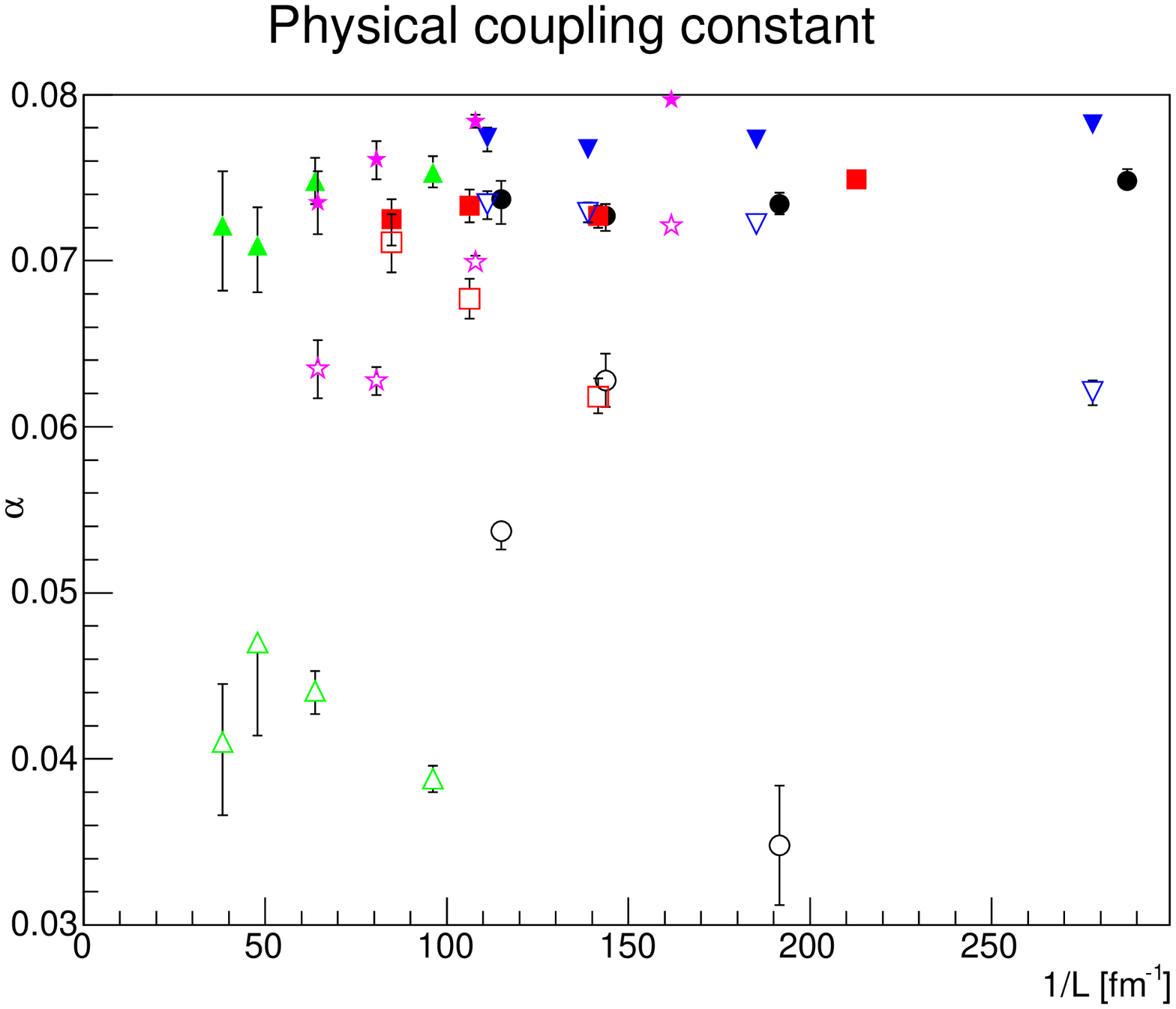}\\
    \includegraphics[width=0.5\textwidth]{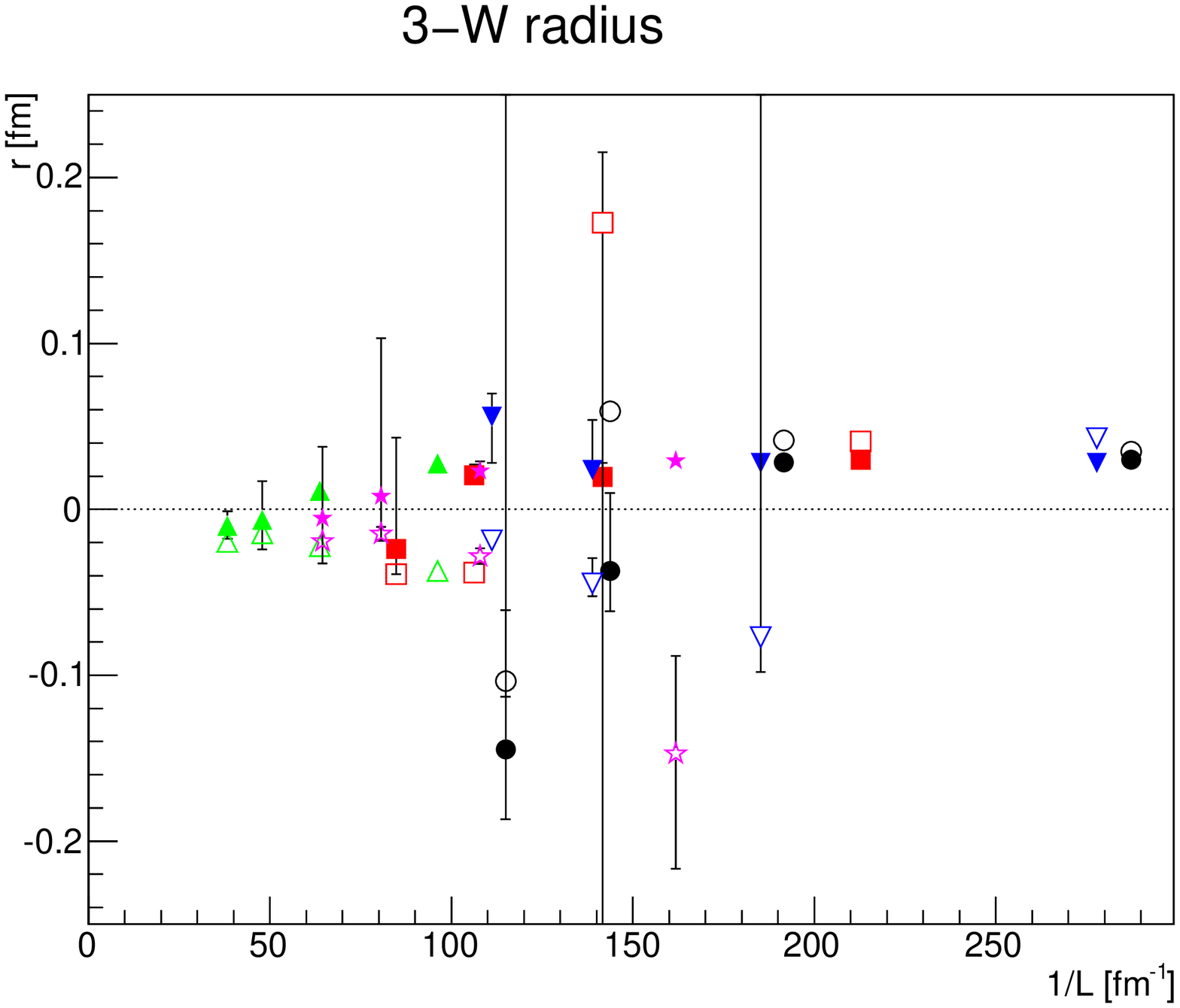}\includegraphics[width=0.5\textwidth]{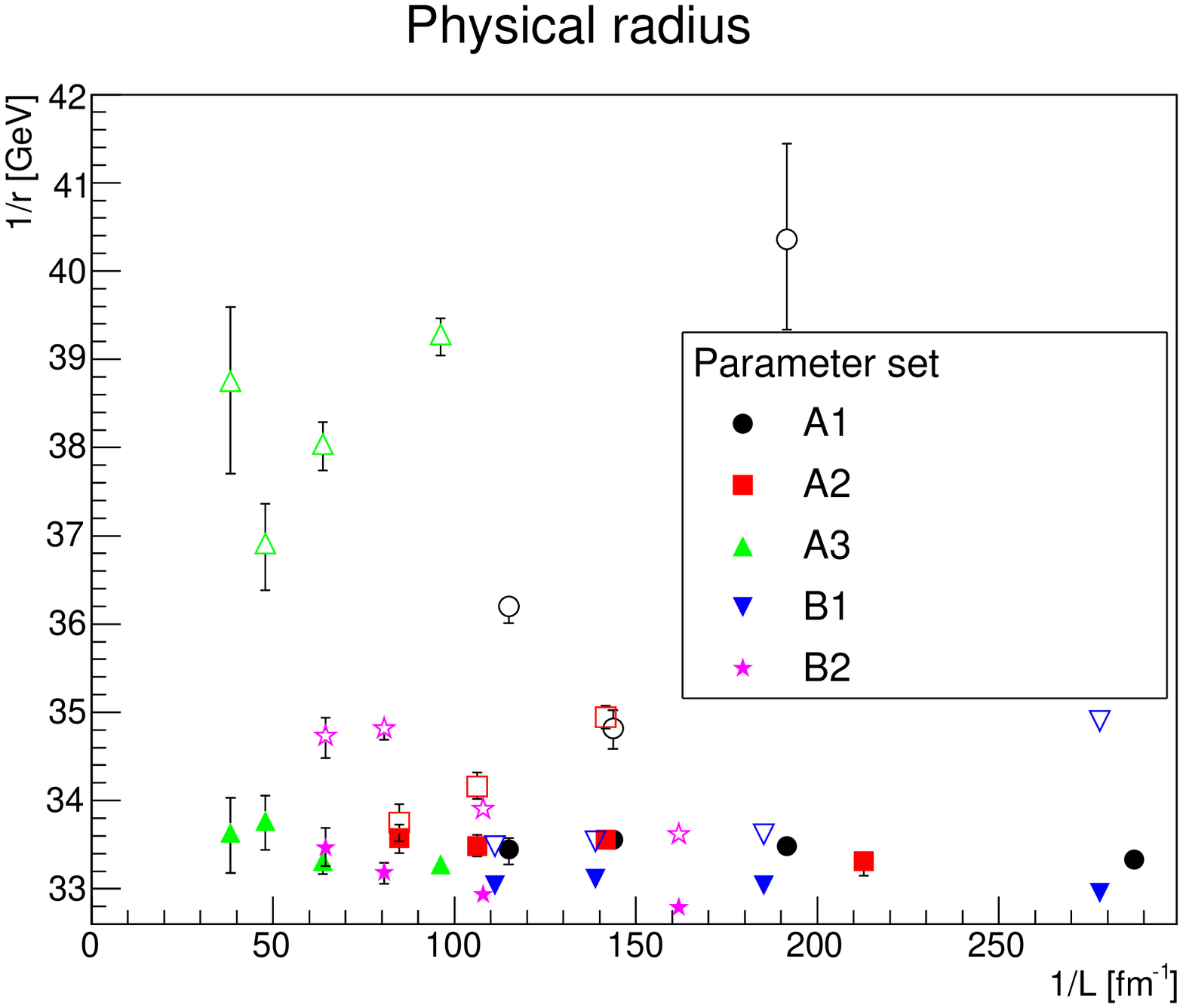}\\
    \caption{The derived reduced effective coupling strength (top panels) and radius (bottom panels). The left panels show the 3-$W$ case and the right panels the physical case. For negative radius squared $\text{sign}(r^2)\sqrt{|r^2|}$ is plotted. For the 3-$W$ case the radius in fm and for the physical case the inverse radius in GeV is plotted, see text. Note the different scales. One point in the physical case and a few points in the 3-W case at small volumes are outside the plot range.}
    \label{fig:radius}
\end{figure}

However, this is dominated by the overall normalization. This is best seen when considering the derived reduced effective coupling $\alpha$ and the radius shown in Figure \ref{fig:radius}, which are defined in Section \ref{ss:ff}. They have been obtained from the fit with errors of $a$ and $b$ propagated.

For the 3-$W$ case the relatively small deviation from a constant leads to large uncertainties in the derived quantities. However, at large volumes the residue, and thus the effective reduced coupling, is small. The radius squared is within errors often compatible with zero. If not, it becomes at large volumes small and negative, indicative of an unphysical particle once more. Interpreting this nonetheless as a radius a value of $1/r\approx 200i$ GeV arises.

In the physical case, the derived quantities depend much less on the parameter set, though they still differ, within statistical error, slightly between sets and between the different momentum configuration. Nonetheless, both the effective reduced coupling and the radius fall into a relatively narrow range of 0.06$\pm$0.02 for the coupling and an inverse radius of 36$\pm$3 GeV, and most values actually cluster even closer together. Thus, $m_wr\approx 2$, which in comparison to, e.g., the proton with a value of 5 shows a rather compact bound state.

This also shows that the effective elementary radius is about five times smaller than the one of the bound state, which perfectly fits with the interpretation above that at large energies the constituents are probed.

\subsection{An experimental setup}\label{ss:exp}

While the quantitative results here should be considered to be rather indications, the qualitative effects are distinct for the gauge-invariant and gauge-dependent case, and can be expected to be robust. Also, taking into account the results on GIPT here, it is not expected that the remainder of the standard model will affect the qualitative outcome, as the remainder is rather affected than source of the difference between physical and unphysical degrees of freedom \cite{Frohlich:1980gj,Frohlich:1981yi,Maas:2017wzi}. Thus, in principle it should be possible to establish the qualitative behavior in experiments. Measuring form factors has a long history already in hadron physics \cite{williams1991nuclear,Dissertori:2003pj,Pacetti:2015iqa}, but there it was easy to prepare all involved particles in an experiment. The situation here is different, as the physical vector bosons inherit the pole structure, and thus width, from the gauge bosons by virtue of \pref{eq:fmsprop}, and they have therefore the observed \cite{pdg} life-time.

However, in principle the measurement techniques developed for atgc \cite{pdg,Baak:2013fwa,Gounaris:1996rz} could be used. For this, it is necessary to reconstruct in an experiment two weak vector bosons\footnote{Of course, alternatively a vector-boson fusion process could be used with a single vector boson in the final state. But for the momentum resolution it would be better to have direct access to two of the vector bosons in the final state.}, say the (physical equivalents to) $W^\pm$. Of course, they will be needed to be reconstructed from their decay products, a highly non-trivial challenge. Their invariant mass $m_{W^+W^-}^2$ must then match the $Z$ mass, showing that they originate from a three-particle interaction. The corresponding cross section needs then to be reconstructed as a function of the four-momenta of the three involved particles. Especially, for a direct comparison this would be needed for off-shell, spacelike momenta, which will complicate the reconstruction procedure. But at least in principle, this should be possible. Alternatively, assuming that \pref{vmd} faithfully captures the time-like domain a comparison to the obtained fits could be made at any momentum configuration.

Thus, the interesting process would be, e.g., at LHC,  $pp\to W^{+*}W^{-*}+X$, where $X$ can be anything, and $m_{W^+W^-}^2=m_{Z}^2$. This puts still the $Z$ on-shell, but currently this seems the best possibility to reconstruct the interesting process. If the four momenta of the $W^{\pm *}$ can be reconstructed, this can be used to reconstruct the form factor at $\Gamma(-m_{Z}^2,Q_+^2,Q_-^2)$ with $(Q_++Q_-)^2=-m_{Z}^2$, where the minus comes from the definition that Euclidean momenta are positive. This can be connected by analytical continuation to the form factor on the lattice or to directly determine the weak radius.

\section{Conclusion}\label{s:con}

Concluding, we have studied for the first time off-shell and interaction properties of the physical vector particles in the weak sector. Fulfilling our primary objective, we do find that they exhibit physical properties, as is required on quite general principles \cite{Seiler:1982pw}. In contrast, the gauge-dependent correlation functions for the $W^\pm$ and $Z$ gauge bosons, which we evaluated on the lattice and therefore automatically include all orders in perturbation theory, show distinctively unphysical features off-shell. This is true both for the particles themselves as well as their interactions. We emphasize here that, despite them being absolutely stable in our simulations, even for unstable particles their behavior is not consistent with physical states. At any rate, their instability is only a parametric effect in the standard model, and not a qualitative one. This underlines the necessity to consider only gauge-invariant composite objects \cite{Banks:1979fi,Frohlich:1980gj,Frohlich:1981yi,Maas:2017wzi} as physical degrees of freedom. The results indicate that the physical vector boson is indeed an extended, though still compact, object.

The features of the physical particles, especially their form factors, should be accessible in experiments, and we outlined a possible setup where this could be done. An actual comparison to experiment may require to access these properties in the time-like domain, which is not possible directly on the lattice.

As an alternative approach we studied as our secondary objective how strongly the physical properties deviate from the analytical predictions of GIPT. We find that we likely will need to go beyond leading order in $v$ in GIPT to do so, especially off-shell and at low energies. This gives us an alternative route for experimental tests, which will also allow to include the remainder of the standard model, which is currently not possible in lattice simulations for various reasons \cite{Maas:2017wzi}. Further, analytic calculations in GIPT should establish how far this needs to be performed by comparing them to the results here.

Altogether the present results not only lay out the next steps to experimentally check the necessity of a fully gauge-invariant prescription of (electro)weak physics, but also provides the first steps in this direction. Even though the lattice results still require substantial improvements in terms of systematics and statistics to reach a quantitative level, rather than the qualitative to semi-quantitative level here.\\

\no{\bf Acknowledgments}\\

We are grateful to Robert Sch\"ofbeck for helpful discussions on Section \ref{ss:exp} and to Wolfgang Schweiger and Helios Sanchis-Alepuz for discussions on form factors. Simulations were performed on the HPC clusters at the University of Graz. The authors are grateful to the HPC teams for the very good performance of the clusters.

\bibliographystyle{bibstyle}
\bibliography{bib}


\end{document}